\documentclass[10pt, journal, final]{IEEEtran}
\usepackage{gensymb}
\usepackage{dsfont}
\usepackage{amsmath,amssymb,amsfonts,amsthm}
\usepackage{epsfig}
\usepackage{cite}
\usepackage{hhline}
\usepackage{pgfplots}
\usepackage{multirow}
\usepackage{xcolor}
\usepackage{makecell}
\usepackage{subcaption}
\usepackage{capt-of}
\usepackage[labelformat=simple]{subcaption}

\usepackage{enumerate}
\usepackage{enumitem}
\usepackage{color}
\usepackage{graphicx}
\usepackage{algorithmic}
\usepackage[ruled]{algorithm}
\usepackage{booktabs}
\usepackage{bm}
\usepackage{tikz}
\usetikzlibrary{automata,arrows,positioning,calc,matrix,decorations.markings,decorations.pathreplacing}

\makeatletter
\newcommand{\vast}{\bBigg@{3}}
\newcommand{\Vast}{\bBigg@{4}}
\makeatother
\renewcommand{\IEEEQED}{\IEEEQEDopen}

\showoutput
\begin{document}

{
\begin{titlepage}
        \vspace*{1.5cm}
\noindent This article has been accepted for publication in the November 2016 issue of
IEEE Journal on Selected Areas in Communications, Special Issue on ``Spectrum Sharing and Aggregation for Future Wireless Networks''.

\vspace{6mm}

\noindent PLEASE CITE THE PAPER AS FOLLOWS:

\vspace{3mm}

\noindent \textbf{Plain text:} \newline
H. Shokri-Ghadikolaei, F. Boccardi, C. Fischione, G. Fodor, and M. Zorzi, ``Spectrum Sharing in mmWave Cellular Networks via Cell Association, Coordination, and Beamforming,'' \emph{IEEE J. Sel. Areas Commun.}, Nov. 2016.

\vspace{3mm}

\noindent \textbf{BibTex:}
\newline @Article\{Shokri2016Spectrum,
\newline  Title                    = \{Spectrum Sharing in \{mmWave\} Cellular Networks via Cell Association, Coordination, and Beamforming\},
\newline  Author                   = \{H. Shokri-Ghadikolaei and F. Boccardi and C. Fischione and G. Fodor and M. Zorzi\},
\newline  Journal=\{IEEE J. Sel. Areas Commun.\},
\newline  Year                     = \{2016\},
\newline  Month                    = \{Nov.\},
\}

\end{titlepage}
}

\cleardoublepage
\title{Spectrum Sharing in mmWave Cellular Networks via Cell Association, Coordination, and Beamforming}

\author{Hossein Shokri-Ghadikolaei,~\IEEEmembership{Student Member,~IEEE,} Federico Boccardi,~\IEEEmembership{Senior Member,~IEEE,} \\Carlo Fischione,~\IEEEmembership{Member,~IEEE,} G\'{a}bor Fodor,~\IEEEmembership{Senior Member,~IEEE,} and Michele Zorzi,~\IEEEmembership{Fellow,~IEEE}
\thanks{H. Shokri-Ghadikolaei and C. Fischione are with KTH Royal Institute of Technology, Stockholm, Sweden (email: hshokri, carlofi@kth.se).}
\thanks{F. Boccardi is with Ofcom, UK (e-mail: federico.boccardi@ieee.org). However, the views
expressed in this research paper are his own and do not reflect those of his employer.}
\thanks{G. Fodor is with the School of Electrical Engineering, KTH Royal Institute of Technology, Stockholm, Sweden, and also with Ericsson Research, Kista, Sweden (e-mail: gaborf@kth.se).}
\thanks{M. Zorzi is with the Department of Information Engineering, University of Padova, Padova, Italy (e-mail: zorzi@dei.unipd.it).}
\thanks{The work of H.~Shokri-Ghadikolaei and C.~Fischione was supported by the Swedish Research Council under the project ``In-Network Optimization".
The work of G. Fodor was supported in part by the Swedish Foundation for Strategic Research through the Strategic Mobility Matthew Project under Grant SM13-0008 and by the joint Ericsson Research--ACCESS Linnaeus Center (KTH) project HARALD.
The work of M.~Zorzi has been partially supported by NYU-Wireless.}
\thanks{G. Fodor would like to thank Dr.~J.~Kronander and Dr.~M.~Prytz, both at Ericsson Research, for their
valuable comments on the manuscript.}}


\newtheorem{theorem}{Theorem}
\newtheorem{defin}{Definition}
\newtheorem{prop}{Proposition}
\newtheorem{lemma}{Lemma}
\newtheorem{corollary}{Corollary}
\newtheorem{alg}{Algorithm}
\newtheorem{remark}{Remark}
\newtheorem{result}{Result}
\newtheorem{conjecture}{Conjecture}
\newtheorem{example}{Example}
\newtheorem{notations}{Notations}
\newtheorem{assumption}{Assumption}

\newcommand{\be}{\begin{equation}}
\newcommand{\ee}{\end{equation}}
\newcommand{\ba}{\begin{array}}
\newcommand{\ea}{\end{array}}
\newcommand{\bea}{\begin{eqnarray}}
\newcommand{\eea}{\end{eqnarray}}
\newcommand{\netsize}[3]{#1x#2~#3$^2$}
\newcommand{\combin}[2]{\ensuremath{ \left( \ba{c} #1 \\ #2 \ea \right) }}
\newcommand{\diag}{{\mbox{diag}}}
\newcommand{\rank}{{\mbox{rank}}}
\newcommand{\dom}{{\mbox{dom{\color{white!100!black}.}}}}
\newcommand{\range}{{\mbox{range{\color{white!100!black}.}}}}
\newcommand{\image}{{\mbox{image{\color{white!100!black}.}}}}
\newcommand{\herm}{^{\mbox{\scriptsize H}}}  
\newcommand{\sherm}{^{\mbox{\tiny H}}}       
\newcommand{\tran}{^{\mbox{\scriptsize T}}}  
\newcommand{\tranIn}{^{\mbox{-\scriptsize T}}}  
\newcommand{\card}{{\mbox{\textbf{card}}}}
\newcommand{\asign}{{\mbox{$\colon\hspace{-2mm}=\hspace{1mm}$}}}
\newcommand{\ssum}[1]{\mathop{ \textstyle{\sum}}_{#1}}

\newcommand{\vbar}{\raisebox{.17ex}{\rule{.04em}{1.35ex}}}
\newcommand{\vbarind}{\raisebox{.01ex}{\rule{.04em}{1.1ex}}}
\newcommand{\D}{\ifmmode {\rm I}\hspace{-.2em}{\rm D} \else ${\rm I}\hspace{-.2em}{\rm D}$ \fi}
\newcommand{\T}{\ifmmode {\rm I}\hspace{-.2em}{\rm T} \else ${\rm I}\hspace{-.2em}{\rm T}$ \fi}
\newcommand{\B}{\ifmmode {\rm I}\hspace{-.2em}{\rm B} \else \mbox{${\rm I}\hspace{-.2em}{\rm B}$} \fi}
\newcommand{\Hil}{\ifmmode {\rm I}\hspace{-.2em}{\rm H} \else \mbox{${\rm I}\hspace{-.2em}{\rm H}$} \fi}
\newcommand{\C}{\ifmmode \hspace{.2em}\vbar\hspace{-.31em}{\rm C} \else \mbox{$\hspace{.2em}\vbar\hspace{-.31em}{\rm C}$} \fi}
\newcommand{\Cind}{\ifmmode \hspace{.2em}\vbarind\hspace{-.25em}{\rm C} \else \mbox{$\hspace{.2em}\vbarind\hspace{-.25em}{\rm C}$} \fi}
\newcommand{\Q}{\ifmmode \hspace{.2em}\vbar\hspace{-.31em}{\rm Q} \else \mbox{$\hspace{.2em}\vbar\hspace{-.31em}{\rm Q}$} \fi}
\newcommand{\Z}{\ifmmode {\rm Z}\hspace{-.28em}{\rm Z} \else ${\rm Z}\hspace{-.38em}{\rm Z}$ \fi}

\newcommand{\sgn}{\mbox {sgn}}
\newcommand{\var}{\mbox {var}}
\newcommand{\E}{\mbox {E}}
\newcommand{\cov}{\mbox {cov}}
\renewcommand{\Re}{\mbox {Re}}
\renewcommand{\Im}{\mbox {Im}}
\newcommand{\cum}{\mbox {cum}}

\renewcommand{\vec}[1]{{\bf{#1}}}     
\newcommand{\vecsc}[1]{\mbox {\boldmath \scriptsize $#1$}}     
\newcommand{\itvec}[1]{\mbox {\boldmath $#1$}}
\newcommand{\itvecsc}[1]{\mbox {\boldmath $\scriptstyle #1$}}
\newcommand{\gvec}[1]{\mbox{\boldmath $#1$}}

\newcommand{\balpha}{\mbox {\boldmath $\alpha$}}
\newcommand{\bbeta}{\mbox {\boldmath $\beta$}}
\newcommand{\bgamma}{\mbox {\boldmath $\gamma$}}
\newcommand{\bdelta}{\mbox {\boldmath $\delta$}}
\newcommand{\bepsilon}{\mbox {\boldmath $\epsilon$}}
\newcommand{\bvarepsilon}{\mbox {\boldmath $\varepsilon$}}
\newcommand{\bzeta}{\mbox {\boldmath $\zeta$}}
\newcommand{\boldeta}{\mbox {\boldmath $\eta$}}
\newcommand{\btheta}{\mbox {\boldmath $\theta$}}
\newcommand{\bvartheta}{\mbox {\boldmath $\vartheta$}}
\newcommand{\biota}{\mbox {\boldmath $\iota$}}
\newcommand{\blambda}{\mbox {\boldmath $\lambda$}}
\newcommand{\bmu}{\mbox {\boldmath $\mu$}}
\newcommand{\bnu}{\mbox {\boldmath $\nu$}}
\newcommand{\bxi}{\mbox {\boldmath $\xi$}}
\newcommand{\bpi}{\mbox {\boldmath $\pi$}}
\newcommand{\bvarpi}{\mbox {\boldmath $\varpi$}}
\newcommand{\brho}{\mbox {\boldmath $\rho$}}
\newcommand{\bvarrho}{\mbox {\boldmath $\varrho$}}
\newcommand{\bsigma}{\mbox {\boldmath $\sigma$}}
\newcommand{\bvarsigma}{\mbox {\boldmath $\varsigma$}}
\newcommand{\btau}{\mbox {\boldmath $\tau$}}
\newcommand{\bupsilon}{\mbox {\boldmath $\upsilon$}}
\newcommand{\bphi}{\mbox {\boldmath $\phi$}}
\newcommand{\bvarphi}{\mbox {\boldmath $\varphi$}}
\newcommand{\bchi}{\mbox {\boldmath $\chi$}}
\newcommand{\bpsi}{\mbox {\boldmath $\psi$}}
\newcommand{\bomega}{\mbox {\boldmath $\omega$}}

\newcommand{\bolda}{\mbox {\boldmath $a$}}
\newcommand{\bb}{\mbox {\boldmath $b$}}
\newcommand{\bc}{\mbox {\boldmath $c$}}
\newcommand{\bd}{\mbox {\boldmath $d$}}
\newcommand{\bolde}{\mbox {\boldmath $e$}}
\newcommand{\boldf}{\mbox {\boldmath $f$}}
\newcommand{\bg}{\mbox {\boldmath $g$}}
\newcommand{\bh}{\mbox {\boldmath $h$}}
\newcommand{\bp}{\mbox {\boldmath $p$}}
\newcommand{\bq}{\mbox {\boldmath $q$}}
\newcommand{\br}{\mbox {\boldmath $r$}}
\newcommand{\bt}{\mbox {\boldmath $t$}}
\newcommand{\bu}{\mbox {\boldmath $u$}}
\newcommand{\bv}{\mbox {\boldmath $v$}}
\newcommand{\bw}{\mbox {\boldmath $w$}}
\newcommand{\bx}{\mbox {\boldmath $x$}}
\newcommand{\by}{\mbox {\boldmath $y$}}
\newcommand{\bz}{\mbox {\boldmath $z$}}

\newenvironment{Ex}
{\begin{adjustwidth}{0.04\linewidth}{0cm}
\begingroup\small
\vspace{-1.0em}
\raisebox{-.2em}{\rule{\linewidth}{0.3pt}}
\begin{example}
}
{
\end{example}
\vspace{-5mm}
\rule{\linewidth}{0.3pt}
\endgroup
\end{adjustwidth}}

\newcommand{\Hossein}[1]{{\textcolor{blue}{\emph{**Hossein: #1**}}}}
\newcommand{\Challenge}[1]{{\textcolor{red}{#1}}}
\newcommand{\NEW}[1]{{\textcolor{blue}{#1}}}


\makeatletter
\let\save@mathaccent\mathaccent
\newcommand*\if@single[3]{%
  \setbox0\hbox{${\mathaccent"0362{#1}}^H$}%
  \setbox2\hbox{${\mathaccent"0362{\kern0pt#1}}^H$}%
  \ifdim\ht0=\ht2 #3\else #2\fi
  }
\newcommand*\rel@kern[1]{\kern#1\dimexpr\macc@kerna}
\newcommand*\widebar[1]{\@ifnextchar^{{\wide@bar{#1}{0}}}{\wide@bar{#1}{1}}}
\newcommand*\wide@bar[2]{\if@single{#1}{\wide@bar@{#1}{#2}{1}}{\wide@bar@{#1}{#2}{2}}}
\newcommand*\wide@bar@[3]{%
  \begingroup
  \def\mathaccent##1##2{%
    \let\mathaccent\save@mathaccent
    \if#32 \let\macc@nucleus\first@char \fi
    \setbox\z@\hbox{$\macc@style{\macc@nucleus}_{}$}%
    \setbox\tw@\hbox{$\macc@style{\macc@nucleus}{}_{}$}%
    \dimen@\wd\tw@
    \advance\dimen@-\wd\z@
    \divide\dimen@ 3
    \@tempdima\wd\tw@
    \advance\@tempdima-\scriptspace
    \divide\@tempdima 10
    \advance\dimen@-\@tempdima
    \ifdim\dimen@>\z@ \dimen@0pt\fi
    \rel@kern{0.6}\kern-\dimen@
    \if#31
      \overline{\rel@kern{-0.6}\kern\dimen@\macc@nucleus\rel@kern{0.4}\kern\dimen@}%
      \advance\dimen@0.4\dimexpr\macc@kerna
      \let\final@kern#2%
      \ifdim\dimen@<\z@ \let\final@kern1\fi
      \if\final@kern1 \kern-\dimen@\fi
    \else
      \overline{\rel@kern{-0.6}\kern\dimen@#1}%
    \fi
  }%
  \macc@depth\@ne
  \let\math@bgroup\@empty \let\math@egroup\macc@set@skewchar
  \mathsurround\z@ \frozen@everymath{\mathgroup\macc@group\relax}%
  \macc@set@skewchar\relax
  \let\mathaccentV\macc@nested@a
  \if#31
    \macc@nested@a\relax111{#1}%
  \else
    \def\gobble@till@marker##1\endmarker{}%
    \futurelet\first@char\gobble@till@marker#1\endmarker
    \ifcat\noexpand\first@char A\else
      \def\first@char{}%
    \fi
    \macc@nested@a\relax111{\first@char}%
  \fi
  \endgroup
}
\makeatother 
\maketitle

\begin{abstract}
This paper investigates the extent to which spectrum sharing in mmWave networks with multiple cellular operators is a viable alternative to traditional dedicated spectrum allocation. Specifically, we develop a general mathematical framework by which to characterize the performance gain that can be obtained when spectrum sharing is used, as a function of the underlying beamforming, operator coordination, bandwidth, and infrastructure sharing scenarios. The framework is based on joint beamforming and cell association optimization, with the objective of maximizing the long-term throughput of the users. Our asymptotic and non-asymptotic performance analyses reveal five key points: (1) spectrum sharing with light on-demand intra- and inter-operator coordination is feasible, especially at higher mmWave frequencies (for example, 73 GHz); (2) directional communications at the user equipments substantially alleviate the potential disadvantages of spectrum sharing (such as higher multiuser interference); (3) large numbers of antenna elements can reduce the need for coordination and simplify the implementation of spectrum sharing; (4) while inter-operator coordination can be neglected in the large-antenna regime, intra-operator coordination can still bring gains by balancing the network load; and (5) critical control signals among base stations, operators, and user equipment should be protected from the adverse effects of spectrum sharing, for example by means of exclusive resource allocation. The results of this paper, and their extensions obtained by relaxing some ideal assumptions, can provide important insights for future standardization and spectrum policy.
\end{abstract}


\section{Introduction}\label{sec: Introduction}
The  scarcity  of  available  sub-6~GHz  spectrum and the significant predicted increase in demand for mobile services have led researchers, both in academia and in industry, to look at millimeter-wave (mmWave) systems, operating at frequencies between 10 and 300~GHz, as a frontier for wireless communication.
Use cases for mmWave networks include backhaul links, mmWave hotspots, and heterogeneous and homogeneous cellular networks~\cite{Rappaport2013Millimeter,Rangan2014Millimeter,Baldemair2015Ultra,Rebatoy2016Potential,shokri2015mmWavecellular,Singh2015TractableModel,di2014stochastic,dehos2014millimeter}.
Meanwhile, preparatory studies for the 2019 World  Radio  Conference are ongoing in order to identify new harmonized bands for mobile services in the mmWave frequency range. Unfortunately,  the  availability  of  spectrum  for mobile  services  presents  limitations  even  at  mmWave  frequencies,  particularly  considering the requirements of other systems that may use these bands in the future, including satellite and  fixed  services \cite{Boccardi2016Spectrum,Guidolin2015InvestigatingGC,Guidolin2015Cooperative,Guidolin2015Study}.  This  is  further  exacerbated  if  we  also  consider  the  need  to  license  mobile bands to multiple operators\footnote{For notation simplicity, in this paper, ``operator''  is synonymous to ``network operator'' and ``service provider.''}, thereby fostering healthy competition in the market. Therefore, it is essential  to  seek  an  optimal  use  of  the  spectrum,  with the  ultimate  goal  of  maximizing benefits for citizens.

\subsection{State of the Art on Spectrum Sharing at mmWave}
Spectrum sharing between multiple operators was recently proposed as a way to allow a more efficient use of the spectrum in mmWave networks. Preliminary studies have shown that the specific  features  of  mmWave  frequencies,  including  the  propagation  characteristics  and  the narrow beams due to directional beamforming, are important enablers for spectrum sharing. In~\cite{Feng2014InterNetwork}, a mechanism that allows two different  IEEE 802.11ad access points to transmit over the same  time/frequency  resources  was  proposed.  This mechanism is  realized  by  introducing  a  new  signaling report, which is broadcast by each access point to establish an interference database that facilitates scheduling  decisions.  A  similar  approach  was  proposed  in~\cite{Li2014Coordniation}  for  mmWave  cellular systems,  with  both  centralized  and  distributed  coordination among operators.  In  the  centralized case,  a  new  architectural  entity  receives  information  about  the  interference  measured  by  each network  and  determines  which  links  cannot  be  scheduled  simultaneously.  In  the  decentralized case,   the   victim  network   sends   a   message   to   the   interfering   network   with   a   proposed coordination  pattern.  The  two  networks  can  further  refine  the  coordination  pattern  via  multiple iterations.

Reference~\cite{niu2015exploiting} investigated the feasibility of sharing the mmWave spectrum between the D2D/cellular and access/backhaul networks and proposed a new MAC layer in order to regulate concurrent transmissions in a centralized manner. Reference~\cite{Shokri2015Transitional} investigated the interference in uncoordinated ad hoc mmWave networks and showed that, under certain conditions, simple scheduling policies with no coordination can be as good as the complicated ones with full coordination. The main reason is the sporadic presence of strong interference that requires on-demand handling. This is in agreement with the recent analysis of the MAC layer of mmWave cellular networks, which shows the need for only on-demand inter-cell interference coordination~\cite{shokri2015mmWavecellular}.
Reference~\cite{Gupta2016Sharing} investigated the feasibility of spectrum sharing in mmWave cellular networks and showed that, under certain conditions such as idealized antenna pattern, spectrum sharing may be beneficial even without any coordination in the entire network. Reference~\cite{Rebatoy2016Potential} showed that infrastructure sharing in mmWave cellular networks is also beneficial and its gain is almost identical to that of spectrum sharing. Reference~\cite{Boccardi2016Spectrum} discussed the architectures and protocols required to make spectrum sharing work in practical mmWave cellular networks and provided preliminary results regarding the importance of coordination. Reference~\cite{rebato2016hybrid} studied the performance of a hybrid spectrum scheme in which exclusive access is used at frequencies in the 20/30~GHz range while spectrum sharing (or even unlicensed spectrum) is used at frequencies around 70~GHz.

\begin{figure}[t]
  \centering
  \includegraphics[width=\columnwidth]{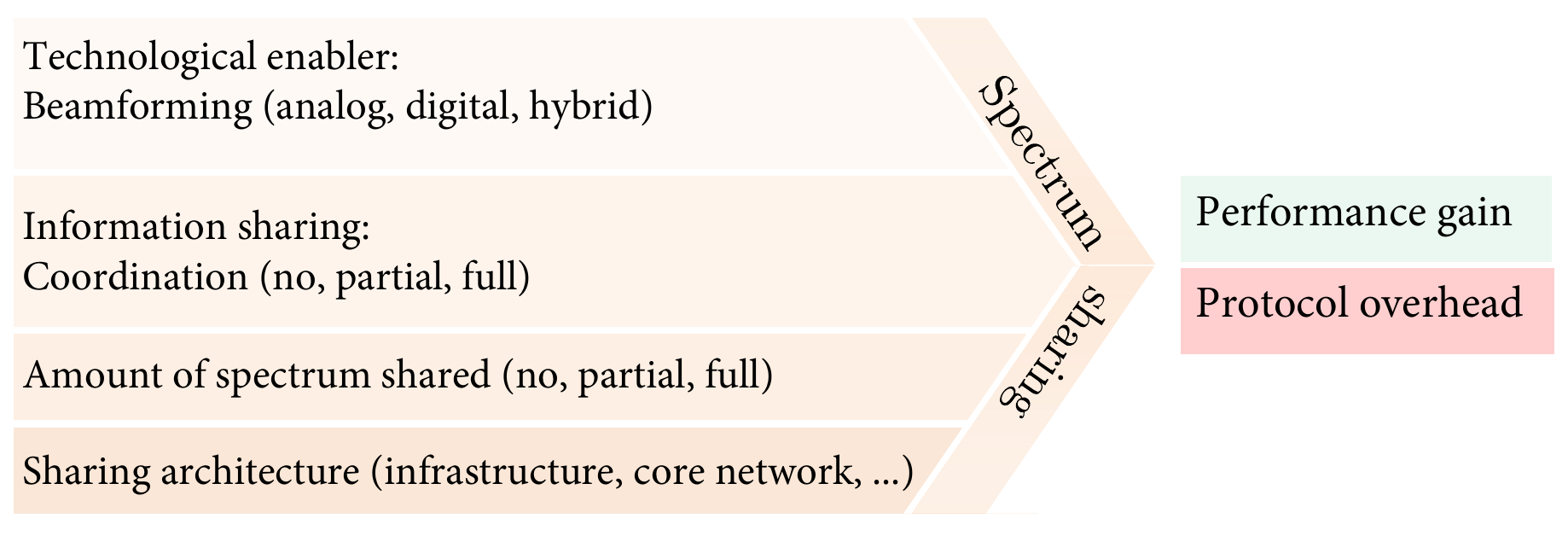}

  \caption{Four aspects have major impacts on the performance of spectrum sharing scenarios, namely the choice of beamforming, the amount of information exchange (coordination), the amount of spectrum shared, and the sharing architecture. This paper focuses on the first three aspects. For a given scenario, there is a tradeoff between performance gain and protocol overhead.}\label{fig: Taxonomy}
\end{figure}
There are four prominent aspects having major impacts on the performance of spectrum sharing schemes: the choice of beamforming, the amount of information exchange (coordination), the amount of spectrum shared, and the sharing architecture (see Fig.~\ref{fig: Taxonomy}).
\begin{itemize}
\item \emph{Beamforming}, both at the transmitter and at the receiver, is the main technological enabler of spectrum sharing in mmWave networks. Digital, analog, and hybrid beamforming are prominent options to realize directional communications. The most appropriate strategy depends on the required flexibility in beam formation, processing and power consumption constraints, cost, and feedback limitations~\cite{venkateswaran2010analog,el2014spatially,Bogale2015Hybrid,Abbas2016Toward}.
\item The \emph{level of coordination}, both within an operator (hereafter called \emph{intra-operator coordination}) and among different operators (hereafter called \emph{inter-operator coordination}), also has a substantial influence on the feasibility of sharing. Coordination includes exchange of channel estimation and precoding vectors.
    Without any coordination (either inter- or intra-operator), spectrum sharing may not even be beneficial in general, especially in traditional sub-6~GHz networks. Full coordination may bring substantial performance gains in terms of throughput and energy, especially because it enables complementary techniques such as joint precoding and load balancing~\cite{Li:15}. From a technical perspective, the challenge is the high overhead of channel estimation and the unaffordable complexity in the core networks in a multi-operator scenario.
\item Another important parameter is the \emph{amount of spectrum shared} among the operators. Increasing the bandwidth (for example, by sharing the spectrum of more operators) may improve the achievable capacity by a prelog factor (that is, a factor outside the $\log$ function), but usually
    reduces the signal-to-interference-plus-noise ratio (SINR), due to higher noise and interference powers. Spectrum sharing is beneficial if the contribution of the prelog factor overweighs the resulting SINR reduction.
\item \emph{Sharing architecture} is another aspect that affects the performance of spectrum sharing. Infrastructure sharing is an example of architecture that increases the gain of spectrum sharing.
\end{itemize}

For a given scenario, the tradeoff between performance gain and protocol overhead determines whether spectrum sharing is beneficial.
In~\cite{Boccardi2016Spectrum}, we discussed various supporting architectures for spectrum sharing in mmWave networks, including interfaces at the radio access networks and at the core network, infrastructure sharing, core network sharing, and spectrum broker. In the present paper, we focus on the first three aspects of Fig.~\ref{fig: Taxonomy}, and leave a detailed investigation of sharing architectures for future works.

\subsection{Contribution of the Present Work}
The motivation for our work is to investigate the gains of beamforming and coordination for spectrum sharing schemes in mmWave networks. To this end, we develop a general mathematical framework based on a joint beamforming design and base station (BS) association, with the objective of maximizing the long-term throughput with fairness guarantees (load balancing). We initially focus on an analog beamforming architecture and formulate a multi-objective optimization problem that finds the optimal association and beamforming when full coordination is available.
We then apply this optimization problem to the cases when either only intra-operator coordination or no coordination is available.
These optimization problems give lower bounds for the performance of spectrum sharing. We then extend this study to a digital beamforming architecture and consider different coordination options, which characterize upper bounds for the performance gain. Our proposed framework makes it possible to incorporate any scenario in terms of the amount of spectrum shared and sharing architecture (see Fig.~\ref{fig: Taxonomy}), and also any spatial distributions of the mobile terminals (UEs)\footnote{According to 3GPP terminology, a mobile terminal is called \emph{user equipment (UE)}.} and the BSs. We also analyze these optimization problems in the asymptotic regimes when the number of antennas, either at the BSs or at the UEs, becomes large. We assess the performance of inter-operator coordination in those regimes.
In addition to the analysis of asymptotic performance, we discuss how limited feedback, imperfect channel state information (CSI) knowledge, and quantized beamforming codebooks affect the sharing performance.

The main conclusions of our work are as follows: (1) spectrum sharing with light on-demand intra- and inter-operator coordination is feasible, especially at higher mmWave frequencies (for example, 73~GHz); (2) directional communications at the UEs substantially alleviate the potential disadvantages of spectrum sharing; (3) larger antenna sizes or equivalently narrower beams reduce the need for coordination and simplify the realization of spectrum sharing; (4) while inter-operator coordination can be neglected in the large antenna regime, intra-operator coordination can still bring gains by balancing the network load; and (5) critical control signals should be protected from the adverse effects of spectrum sharing, for example by means of exclusive resource allocation. We believe that this work provides important contributions to answer the following fundamental questions related to future mmWave networks: for a given amount of contiguous mmWave spectrum for mobile applications, what is the spectrum allocation/access scheme that allows the optimal utilization of the spectrum and the highest benefit to society? How much intra- and inter-operator coordination is required to make spectrum sharing work in mmWave bands? What are the parameters affecting the feasibility of spectrum sharing? How should those parameters be chosen?

The rest of this paper is organized as follows. Section~\ref{sec: system-model} presents the system model used in this paper.
Sections~\ref{sec: pooling-analog-beamforming} and~\ref{sec: pooling-digital-beamforming} formulate optimization problems to study the gains of coordination and beamforming in spectrum sharing schemes in mmWave networks. Section~\ref{sec: assymptotic-performance} analyzes the performance of spectrum sharing in the asymptotic regimes. We present numerical results in Section~\ref{sec: numerical-results}, followed by design insights in Section~\ref{sec: design-insights} and concluding remarks in Section~\ref{sec: concluding-remarks}.

\section{System Model}\label{sec: system-model}
\subsection{Network Model}
We consider the downlink of a multi-operator cellular network with total bandwidth $W$ to be shared among $Z$ operators in the network. Let $\mathcal{B}_z$ be the set of BSs of operator $z$, and $\mathcal{B} = \mathcal{B}_1 \cup \mathcal{B}_1 \cup \ldots  \cup \mathcal{B}_{Z}$ be the set of all BSs in the network. Note that the spatial distribution of the BSs of each operator $z$ and also infrastructure sharing are embedded in $\mathcal{B}_z$. For instance, with no infrastructure sharing, $\{\mathcal{B}_z\}_{z=1}^{Z}$ are disjoint sets. We denote by $\mathcal{U}$ the set of all UEs, by $\mathcal{U}_z$ the set of all UEs of operator $z$, and by $W_z$ the bandwidth of operator $z$.
Without loss of generality, we assume universal frequency reuse within an operator, so all the non-serving BSs of an operator cause interference to every UE of that operator in the downlink.
Table~\ref{table: notations} lists the main symbols used in the paper.
\begin{table}[t]
  \centering
  \caption{Summary of main notations.}\label{table: notations}
{
\renewcommand{\arraystretch}{1.2}
  {
   \begin{tabular}{|@{}c@{}|l|}
\hline
   \textbf{Symbol} & \textbf{Definition} \\ \hline
    $N_{\mathrm{BS}},N_{\mathrm{UE}}$ & Number of antennas at every BS and UE \\ \hline
    $N_{b}$ & Number of UEs that are associated to BS $b$ \\ \hline
    $Z$ & Number of operators \\ \hline
    $N_{b \hskip 0.2mm u}$ & Number of paths between BS $b \in \mathcal{B}$ and UE $u  \in \mathcal{U}$ \\ \hline
    $N_{r}$ & Number of RF chains at each BS \\ \hline
    $\mathcal{U}$ & Set of all UEs of all operators \\ \hline
    $\mathcal{B}_{z}$ & Set of BSs of operator $z$ \\ \hline
    $\mathcal{U}_{z}$ & Set of UEs of operator $z$ \\ \hline
    $W_{z}$ & Bandwidth of operator $z$ \\ \hline
    $\vec{H}_{b \hskip 0.2mm u}$ & Channel matrix between BS $b$ and UE $u$ \\ \hline
    $\vec{H}_{b}$ & Effective channel from the perspective of BS $b$ \\ \hline
    $\vec{H}$ & Effective channel for precoding \\ \hline
    $~\vec{a}_{\mathrm{BS}} (\theta)~$ & Vector response function of the BSs' antenna arrays to $\theta$ \\ \hline
    $~\vec{a}_{\mathrm{UE}} (\theta)~$ & Vector response function of the UEs' antenna arrays to $\theta$ \\ \hline
    $x_{b \hskip 0.2mm u}$ & Binary association variable of UE $u$ to BS $b$\\ \hline
    $\mathcal{A}_{b}$ & Set of UEs that are associated to BS $b$ \\ \hline
    $b$ & An index denoting a tagged BS \\ \hline
    $u$ & An index denoting a tagged UE \\ \hline
    $i$ & An index denoting a BS \\ \hline
    $j$ & An index denoting a UE \\ \hline
    $k,z$ & Indices denoting an operator \\ \hline
\end{tabular}}
}
\end{table}

We denote by $x_{b \hskip 0.2mm u}$ a binary variable that is equal to 1 if UE $u \in \mathcal{U}$ is served by  (or associated to) BS $b \in \mathcal{B}$, and by $N_{b} = \sum_{u \in \mathcal{U}}{x_{b \hskip 0.2mm u}}$ and $\mathcal{A}_{b}$ the number and the set of UEs that are being served by BS $b$, respectively. We also call $N_{b}$ the load of BS $b$. Note that without national roaming, each BS can serve only UEs of the same operator. Namely, $x_{b \hskip 0.2mm u} = 0$ for all $b \in \mathcal{B}_z, u \in \mathcal{U}_k$ where $z\neq k$.
We define the association period as the consecutive coherence intervals over which an association remains unchanged, see Fig.~\ref{fig: SuperframeStructure}. For a given association, beamforming should be recomputed every coherence time, and short-term scheduling should be recalculated after a small number of coherence intervals (e.g., 10~ms in LTE~\cite{sesia2009lte}), whereas association is a long-term process in the sense that it is fixed over many coherence time intervals~\cite{andrews2014overview,Yu2016Distributed,elshaer2016downlink,Athanasiou-etal-2013,Athanasiou2014Auction}. In this paper, we investigate the performance of optimal association; i.e., we find the optimal $\mathcal{A}_{b}$ for all BSs and all operators. Using these associations, the BSs and UEs recalculate their beamforming vectors every coherence time.
For short-term scheduling, we ensure that each BS can serve all its UEs simultaneously (by assuming a sufficiently large number of RF chains at each BS). By this assumption, we avoid the interplay between the short-term resource allocation problem and the association problem, which should be handled at different time scales. This simplification leads to simpler mathematical formulations, whose extension to the general case is left for future work.

\begin{figure}[t]
  \centering
  \includegraphics[width=0.95\columnwidth]{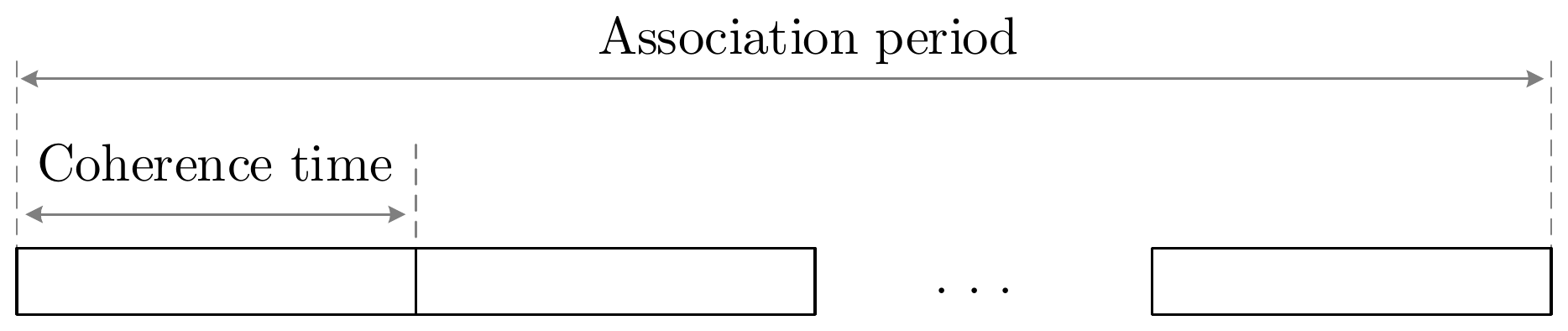}

  \caption{An association period. Beamforming vectors are fixed only for one coherence time, and should be recomputed afterward. Association is fixed over a block of many coherence time intervals, denoted as association period.
  }\label{fig: SuperframeStructure}
\end{figure}

\subsection{Antenna and Channel Model}
We consider a half wavelength uniform linear array (ULA) of $N_{\mathrm{BS}}$ antenna elements for all BSs and a ULA of $N_{\mathrm{UE}}$ antennas for all UEs, albeit our mathematical framework can be easily extended to other antenna models. We consider a narrowband cluster channel model~\cite{el2014spatially}, with a random number of paths between every BS and every UE. This model can be transformed into the well-known virtual channel model of~\cite{sayeed2002deconstructing}. Let $N_{b \hskip 0.2mm u}$ be the number of paths between BS $b \in \mathcal{B}$ and UE $u  \in \mathcal{U}$, and $g_{b \hskip 0.2mm u \hskip 0.2mm n}$ be the complex gain of the $n$-th path that includes both path loss and small scale fading. In particular, $g_{b \hskip 0.2mm u \hskip 0.2mm n}$ is a zero-mean complex Normal random variable with $\mathbb{E}[|g_{b \hskip 0.2mm u \hskip 0.2mm n}|^2] = L_{b \hskip 0.2mm u}$ for $n=1,2,\ldots,N_{b \hskip 0.2mm u}$, where $L_{b \hskip 0.2mm u}$ in the path loss between BS $b$ and UE $u$ and is the product of a constant attenuation, a distance dependent attenuation, and a large scale Lognormal fading~\cite{Akdeniz2014MillimeterWave}.
The channel matrix between BS $b$ and UE $u$ is given by
\begin{equation}\label{eq: channel-matrix}
\vec{H}_{b \hskip 0.2mm u} = \sqrt{\frac{N_{\mathrm{BS}}N_{\mathrm{UE}}}{N_{b \hskip 0.2mm u}}}\sum\limits_{n=1}^{N_{b \hskip 0.2mm u}}{g_{b \hskip 0.2mm u \hskip 0.2mm n} \, \vec{a}_{\mathrm{UE}}\left(\theta_{b \hskip 0.2mm u \hskip 0.2mm n}^{\mathrm{UE}}\right) \vec{a}_{\mathrm{BS}}^{*} \left(\theta_{b \hskip 0.2mm u \hskip 0.2mm n}^{\mathrm{BS}}\right)} ,
\end{equation}
where $\vec{a}_{\mathrm{BS}} \in \mathds{C}^{N_{\mathrm{BS}}}$ and $\vec{a}_{\mathrm{UE}} \in \mathds{C}^{N_{\mathrm{UE}}}$ are the vector response functions of the BSs' and UEs' antenna arrays to the angles of arrival and departure (AoAs and AoDs), $\theta_{b \hskip 0.2mm u \hskip 0.2mm n}^{\mathrm{BS}}$ is the AoD of the $n$-th path, $\theta_{b \hskip 0.2mm u \hskip 0.2mm n}^{\mathrm{UE}}$ is the AoA of the $n$-th path, and $(\cdot)^{*}$ is the conjugate transpose operator.
For a ULA with half wavelength antenna spacing at the BS, we have
\begin{equation}\label{eq: ULA-antenna-response}
\vec{a}_{\mathrm{BS}}\left(\theta\right) = \frac{1}{\sqrt{N_{\mathrm{BS}}}}\left[1, e^{-j \pi \sin(\theta)}, \ldots, e^{-j (N_{\mathrm{BS}}-1)\pi\sin(\theta)}\right]^{*} \:.
\end{equation}
The parameters of the channel model depend both on the carrier frequency and on being in line-of-sight (LoS) or non-LoS conditions and are given in~\cite[Table~I]{Akdeniz2014MillimeterWave}. The AoAs and AoDs depend on the spatial distribution of the BSs and the UEs, and on the scattering in the environment. For instance, if they follow independent homogenous Poisson point processes, $\theta_{b \hskip 0.2mm u \hskip 0.2mm n}^{\mathrm{UE}}$ and $\theta_{b \hskip 0.2mm u \hskip 0.2mm n}^{\mathrm{BS}}$ (for all $b$, $u$, and $n$) will be independent random variables with uniform distribution in~$[0,2\pi]$. The path loss, AoAs, and AoDs depend on the second-order statistics of the wireless channel, which remains unchanged as long as the scattering geometry relative to the corresponding user remains unchanged. Even if the user changes its position by some meters, the channel second order statistics remain unchanged~\cite{rappaport2002wireless}.


\subsection{Beamforming Model}
To reduce UE power consumption while providing sufficiently good performance, we consider an analog combiner using phase shifters at the UE side (only one RF chain per UE). With these phase shifters, each UE can only change its antenna boresight based on a predefined codebook $\mathcal{W}^{\mathrm{UE}}$, whose cardinality mainly depends on the feedback size. For example, 4-bit feedback allows 16 different vectors in the codebook. Supposing that UE $u$ will be served by BS $b$, its combining vector is $\vec{w}_{b \hskip 0.2mm u}^{\mathrm{UE}} \in \mathbb{C}^{N_{\mathrm{UE}}}$.
Let $\vec{W}_{b}^{\mathrm{BS}} \in \mathbb{C}^{N_{\mathrm{BS}}\times N_{b}}$ be the precoding matrix at BS $b$ whose $u$-th column $\vec{w}_{b \hskip 0.2mm u}^{\mathrm{BS}} \in \mathds{C}^{N_{\mathrm{BS}}}$ is the precoding vector for UE $u$.
In the following, we introduce the precoding strategies considered in the next sections. In short, all the UEs always use an analog combiner, whereas the precoder at the BSs can be either analog or digital.
\subsubsection{Analog Precoding}
We assume $N_r $ RF chains at each BS, where $1 \leq N_{r} \ll N_{\mathrm{BS}}$.\footnote{The difference with respect to the hybrid precoder is the lack of digital precoder. In other words, signals of individual RF chains will not be jointly processed here, giving a lower bound for the achievable performance.} Each RF chain can serve only one UE at a time; however, a BS can serve multiple UEs at the same time by exploiting different RF chains. To avoid unnecessary complications due to short-term scheduling, we assume that all the UEs will always be served by their BSs. To ensure this, we put condition ${N_{b} \leq N_r}$ for all BSs and all operators in the next sections.

We assume that the total transmission power $p$ will be divided equally among all active RF chains, so the transmission power of BS $b$ toward its individual UEs is $p/N_b$, where $N_b$ is the BS load (number of its active RF chains).
The received power of UE $u$ from BS $b$ is ${p |\hspace{-0.4mm}\left(\vec{w}_{b \hskip 0.2mm u}^{\mathrm{UE}}\right)^{*} \vec{H}_{b \hskip 0.2mm u} \vec{w}_{b \hskip 0.2mm u}^{\mathrm{BS}}|^2/N_{b}}$. The RF precoding and combining vectors try to maximize the received power. Formally, they are the solution of the following optimization problem:
\begin{subequations}\label{eq: AnalogBeamforming}
\begin{alignat}{3}
\label{eq: analog-beamforming-objective}
&\underset{\vec{w}_{b \hskip 0.2mm u}^{\mathrm{UE}},~\vec{w}_{b \hskip 0.2mm u}^{\mathrm{BS}}}{\text{maximize}} \hspace{3mm} && \left| \left(\vec{w}_{b \hskip 0.2mm u}^{\mathrm{UE}}\right)^{*} \, \vec{H}_{b \hskip 0.2mm u} \, \vec{w}_{b \hskip 0.2mm u}^{\mathrm{BS}} \right|^2 \:,
\\
\label{eq: analog-beamforming-C1}
&{\text{subject to}} && \vec{w}_{b u}^{\mathrm{UE}} \in \mathcal{W}^{\mathrm{UE}}\:,
\\
\label{eq: analog-beamforming-C2}
& &&
\vec{w}_{b u}^{\mathrm{BS}} \in \mathcal{W}^{\mathrm{BS}} \:,
\end{alignat}
\end{subequations}
where $\mathcal{W}^{\mathrm{BS}}$ and $\mathcal{W}^{\mathrm{UE}}$ are the precoding codebooks of the BSs and the UEs, respectively. For a given channel matrix, solving optimization problem~\eqref{eq: AnalogBeamforming} requires, in the worst case, an exhaustive search over $\big|\mathcal{W}^{\mathrm{UE}}\big|\big|\mathcal{W}^{\mathrm{BS}}\big|$ choices of the BS precoders and UE combiners.
The solution of~\eqref{eq: AnalogBeamforming} may vary in every coherence time. However, as we show in Section~\ref{sec: assymptotic-performance}, asymptotically, the optimal analog precoder and combiner depend only on the second order statistics of the channel, as also suggested in~\cite{Nam2014Joint,Bogale2015Hybrid}. This property allows asymptotic performance analysis for spectrum sharing.


\subsubsection{Digital Precoding}
Here, we assume $N_r = N_{\mathrm{BS}}$ RF chains at all BSs. Again, we allow a total transmission power $p$ at each BS. As the beamforming and association problems should be handled at different time scales, we optimize the user association problem in Section~\ref{sec: pooling-digital-beamforming} given a reasonable precoding scheme.
First we note that the transmitted symbols of BS $b$ are $\vec{s}_{b} = \sqrt{\lambda_{b}} \vec{W}_{b}^{\mathrm{BS}}\vec{d}_{b}$, where ${\vec{d}_{b} \in \mathbb{C}^{N_{b}} \sim \mathcal{CN}(\vec{0},\vec{I}_{N_{b}})}$ are the data symbols for the $N_{b}$ UEs of this cell. The parameter $\lambda_{b}$ normalizes the average transmit power of the BS to $p$, namely
\begin{equation}
\lambda_{b} = \frac{p}{\mathbb{E}\left[\mathrm{tr} \, \vec{W}_{b}^{\mathrm{BS}} \left(\vec{W}_{b}^{\mathrm{BS}}\right)^{*}\right]} \:.
\end{equation}
The precoding matrix depends on the precoding scheme and also on the amount of information available at the BS. In this paper, we consider two digital precoding strategies of practical interest: maximum ratio transmission (MRT) and regularized zero forcing (RZF). Define $\overline{\vec{H}}$ as the effective channel that contains the effect of analog combiners of some of the UEs; see Section~\ref{sec: pooling-digital-beamforming} for the formal definition of $\overline{\vec{H}}$.
Given $\overline{\vec{H}}$, the combining vectors are the solution of the following optimization problem:
\begin{subequations}\label{eq: DigitalBeamforming}
\begin{alignat}{3}
\label{eq: digital-beamforming-objective}
&\underset{\vec{w}_{b \hskip 0.2mm u}^{\mathrm{UE}}}{\text{maximize}} \hspace{3mm} && \left| \left(\vec{w}_{b \hskip 0.2mm u}^{\mathrm{UE}}\right)^{*} \, \vec{H}_{b \hskip 0.2mm u} \, \vec{w}_{b \hskip 0.2mm u}^{\mathrm{BS}} \right|^2 \:,
\\
\label{eq: digital-beamforming-C1}
&{\text{subject to}} && \vec{w}_{b \hskip 0.2mm u}^{\mathrm{UE}} \in \mathcal{W}^{\mathrm{UE}}\:,
\\
\label{eq: digital-beamforming-C2}
& &&
\vec{w}_{b \hskip 0.2mm u}^{\mathrm{BS}} =
\begin{cases}
 \left[\overline{\vec{H}}^{*}\right]_{u}, & \mbox{MRT} \\
 \left[\left(\overline{\vec{H}} + c\vec{I}\right)^{\dag}\right]_{u}, & \mbox{RZF} \:,
 \end{cases}
\end{alignat}
\end{subequations}
where $c$ is an arbitrary positive number, $\vec{I}$ is the identity matrix of the same size as $\overline{\vec{H}}$, $[\vec{A}]_u$ is the $u$-th column of matrix $\vec{A}$, and $(\cdot)^{\dag}$ is the pseudo-inverse operation.
In Section~\ref{sec: pooling-digital-beamforming}, we characterize $\overline{\vec{H}}$ based on the information available at each BS.

\subsection{Methodology}\label{sec: methodology}
To investigate the gains of beamforming and coordination for spectrum sharing schemes in mmWave networks, we note that although the information theoretic foundations of wideband communications (implying the SNR approaching zero) have been the topic of intensive research (see, e.g.,~\cite{verdu2002spectral,Dana2007Information,Raghavan2007Capacity}), their applications in practical scenarios are less understood. The problem becomes even more challenging when we take spectrum sharing and various coordination schemes into account. Our approach using a network optimization framework can alleviate such problems.

In the following, we formulate the relevant optimization problems to optimize the performance of spectrum sharing for different beamforming and coordination strategies. The comparison of the solutions of these optimization problems, though not in closed-form, enables us to quantify  the gains of different coordination levels (inter-operator, intra-operator, or uncoordinated) with different beamforming schemes (analog or digital). Pursuing closed-form results is an interesting topic for future work.

\section{Spectrum Sharing with Analog Beamforming}\label{sec: pooling-analog-beamforming}
In this section, we evaluate the gain of spectrum sharing with analog precoding at the BSs when considering different coordination scenarios.


\subsection{SINR and Rate Models}
The received power at each UE $u \in \mathcal{U}_z$ when the serving BS is $b \in \mathcal{B}_z$ is the summation of five components: desired power $P$, intra-cell interference $I_1$, inter-cell interference $I_2$, inter-operator interference $I_3$, and noise power spectral density $\sigma^2$. $I_1$ corresponds to the signals transmitted to other UEs by the same BS. $I_2$ corresponds to the interference from the signals transmitted by other BSs of the same network operator. $I_3$ corresponds to the interference from the signals transmitted by all BSs of other operators $\mathcal{B}\setminus \mathcal{B}_z$ toward their own UEs.

We first note that the received power of UE $u$ from BS $b$ is ${p |\hspace{-0.4mm}\left(\vec{w}_{b \hskip 0.2mm u}^{\mathrm{UE}}\right)^{*} \vec{H}_{b \hskip 0.2mm u} \vec{w}_{b \hskip 0.2mm u}^{\mathrm{BS}}|^2/N_{b}}$. To calculate the interference terms, recall the definitions of the binary association variables $x_{ij}$, the set of associated UEs $\mathcal{A}_{i}$, and the load of a BS $N_{i}$, for each BS $i$ and UE $j$. It follows that
\begin{align}\label{eq: intra-cell-interference}
I_1 & = \sum\limits_{j \in \mathcal{A}_{b} \setminus\{u\}} \frac{p}{N_{b}}\left|\left(\vec{w}_{b u}^{\mathrm{UE}}\right)^{*} \vec{H}_{b u} \vec{w}_{b j}^{\mathrm{BS}}\right|^2 \nonumber \\
& = p\sum\limits_{j \in \mathcal{U}_z \setminus\{u\}} \frac{x_{bj}}{\sum_{m \in \mathcal{U}_z}x_{bm}}\left|\left(\vec{w}_{b u}^{\mathrm{UE}}\right)^{*} \vec{H}_{b u} \vec{w}_{b j}^{\mathrm{BS}}\right|^2 \:,
\end{align}
where $p/N_b$ is the transmit power of BS $b$ toward UE $u$. Similarly, we have
\begin{align}\label{eq: inter-cell-interference}
I_2 & = p \sum\limits_{i \in \mathcal{B}_z \setminus \{b\}} \, \sum\limits_{j \in \mathcal{U}_z} \frac{x_{ij}}{\sum_{m \in \mathcal{U}_z}x_{im}}\left|\left(\vec{w}_{b u}^{\mathrm{UE}}\right)^{*} \vec{H}_{i u} \vec{w}_{i j}^{\mathrm{BS}}\right|^2.
\end{align}

For each UE $u$, inter-operator interference $I_3$ depends on the set of operators (and BSs) that share the same bandwidth. Without loss of generality of the developed mathematical framework, we consider only a full bandwidth sharing scenario, namely the total bandwidth $W$ is reused by all operators, i.e., $W_z = W$. With universal frequency reuse, each UE $u$ receives interference from all BSs of all operators, and the inter-operator interference can be expressed as
\begin{equation}\label{eq: inter-operator-interference}
I_3 = p \sum\limits_{k=1 \hfill \atop k \neq z \hfill}^{Z} \, \sum\limits_{i \in \mathcal{B}_k} \, \sum\limits_{j \in \mathcal{U}_k} \frac{x_{ij}}{\sum_{j \in \mathcal{U}_k}x_{i j}}\left|\left(\vec{w}_{b \hskip 0.2mm u}^{\mathrm{UE}}\right)^{*} \vec{H}_{i u} \vec{w}_{ij}^{\mathrm{BS}}\right|^2 \:.
\end{equation}
Using the special characteristics of mmWave networks, such as high penetration loss and directional communications, interference components~\eqref{eq: intra-cell-interference}--\eqref{eq: inter-operator-interference} can be substantially simplified for mathematical tractability with almost negligible loss in the accuracy of the SINR model~\cite{Shokri2016OntheAccuracy}. In this work, we keep the general form of the interference terms, though their simplified versions can also be substituted in the following expressions.

The average rate that UE $u$ can get from BS $b$ conditioned on the network topology is
\begin{equation}\label{eq: average-rate}
r_{b \hskip 0.2mm u} = \mathbb{E} \left[W\log\left(1 + \frac{P}{I_1 + I_2 + I_3 + W\sigma^2}\right) \right] \:,
\end{equation}
where the expectation is over all random channel gains. The long-term rate that UE $u$ will receive from all BSs of operator $z$ is equal to ${r_{u} = \sum_{b \in \mathcal{B}_z}{x_{b \hskip 0.2mm u}r_{b \hskip 0.2mm u}}}$.
Sharing the spectrum increases the bandwidth available to each operator (with a prelog contribution to the rate in high SINR regimes); however it also increases the interference power. We collect all control variables $x_{b \hskip 0.2mm u}$ in matrix $\vec{X}_z$ for all operators $1\leq z \leq Z$, and let ${\vec{X} = \diag (\vec{X}_1, \vec{X}_2, \ldots, \vec{X}_{Z})}$. For each operator $z$, the intra- and inter-cell interference terms, $I_1$ and $I_2$, are only functions of $\vec{X}_z$. However, the inter-operator interference $I_3$ is a function of the association solutions of other operators $\{\vec{X}_i\}_{i=1,i\neq z}^{Z}$. Therefore, $r_{u}$ depends on the association solutions of all operators $\vec{X}$.

Note that the association solution remains unchanged in the calculation of~\eqref{eq: average-rate}. This solution might not be optimal over some coherence time intervals; however, it leads to the maximum average rate in the long run. We note the following:

\begin{remark}
Equation~\eqref{eq: average-rate} provides a lower bound for the average rate that BS $b$ gives to UE $u$. After finding the association solution based on this objective function, intra-operator and inter-operator coordination in designing time-frequency scheduling can further reduce $I_2$ and $I_3$, respectively. These short-term adaptations statistically improve the rate of individual UEs.
\end{remark}

\subsection{Optimal Association with Sharing and Full Coordination}\label{sec: optimal-association-full-coordination-analog}
To ensure both high network throughput and some level of fairness among individual UEs, we consider proportional fairness as the network utility of each operator. In particular, the objective function of operator $z$ is
\begin{equation}\label{eq: ObjectiveFunction1}
f_z (\vec{X}) = \sum\limits_{u \in \mathcal{U}_z}\,\log r_u = \sum\limits_{u \in \mathcal{U}_z}\,{\log \left(\sum\limits_{b \in \mathcal{B}_z} \, x_{b \hskip 0.2mm u}r_{b \hskip 0.2mm u}\right)} \:.
\end{equation}
Given that the beamforming is recomputed in every coherence time based on~\eqref{eq: AnalogBeamforming}, we can formulate the following multi-objective optimization problem to find the optimal association:
\begin{subequations}\label{eq: long-term-resource-analog}
\begin{alignat}{3}
\mathcal{P}_1\hspace{-0.8mm}:~\hspace{-1.3mm} & \underset{\vec{X}}{\text{maximize}} \hspace{1.6mm} && \left[f_1 (\vec{X}), f_2 (\vec{X}), \ldots,f_{Z} (\vec{X})\right] \:,
\\
& {\text{subject to}} && \mbox{beamforming design~\eqref{eq: AnalogBeamforming}}\:,\\
\label{eq: const-unique-association}
& && \hspace{-2mm}
\sum\limits_{b \in \mathcal{B}_z} x_{b \hskip 0.2mm u} = 1 \:, \hspace{1.5mm} \forall u \in \mathcal{U}_z, 1\leq z \leq Z\:,
\\
\label{eq: cell-size-p1}
& && \hspace{-2mm}\sum\limits_{u \in \mathcal{U}_z} x_{b \hskip 0.2mm u} \leq N_{r} \:, \hspace{1.5mm} \forall b \in \mathcal{B}_z, 1\leq z \leq Z \, \\
\label{eq: binary-variables}
& && \hspace{-2mm}
x_{b \hskip 0.2mm u} \in \{0,1 \} \:, \forall b \in \mathcal{B}, u \in \mathcal{U} \:, \\
\label{eq: binary-variables-11}
& &&\hspace{-2mm}
x_{b \hskip 0.2mm u}=0 \:, \forall b \in \mathcal{B}_k, u \in \mathcal{U}_z, k\neq z, 1 \leq z,k \leq Z.
\end{alignat}
\end{subequations}
Constraint~\eqref{eq: const-unique-association} guarantees association of each UE to only one BS, mitigating joint scheduling requirements among BSs. Constraint~\eqref{eq: cell-size-p1} ensures that $N_{b} \leq N_{r}$, so all $N_{b}$ UEs that are associated to BS $b$ can be served together with MU-MIMO. If $N_{b} < N_{r}$, some RF chains will be off, and the BS gives higher transmit power to the active RF chains. Constraint~\eqref{eq: binary-variables-11} ensures that the UEs of operator $z$ can be only served by BSs of the same operator (no national roaming). Notice that optimization problem~\eqref{eq: long-term-resource-analog} can be easily extended to allow national roaming by just removing~\eqref{eq: binary-variables-11} and making simple modifications of~\eqref{eq: const-unique-association}--\eqref{eq: cell-size-p1}.

Finding the Pareto optimal solution of $\mathcal{P}_1$ directly from the optimization problem is difficult. The general approach to address multi-objective optimization problems is to transform them into a scalar problem (scalarization). Among all the scalarization approaches, the weighted Tchebycheff method can provide the complete Pareto optimal solutions with low computational complexity and is effective even for non-convex objective functions~\cite{Marler2004Survey}. Therefore, we adopt it as the scalarization method in this paper. Note that the main aim of this paper is to understand the feasibility of spectrum sharing in mmWave networks, and efficient solution methods for $\mathcal{P}_1$ are left for future work. In the following subsections, we pose other variations of optimization problem $\mathcal{P}_1$ when we have different coordination levels and beamforming schemes.
By analyzing the solutions of these optimization problems, we characterize the gains of coordination and beamforming for spectrum sharing schemes in mmWave cellular networks.

The solution of $\mathcal{P}_1$ provides long-term resource allocation policies for the network. This solution is valid as long as the input of the optimization problem, i.e., the network topology, remains unchanged. Once either a UE loses its connection (for instance, due to a temporary obstacle), or the channel statistics changes (for instance, due to mobility) optimization problem $\mathcal{P}_1$ has to be re-executed.

\begin{remark}[Operating Frequency]
Changing the frequency band affects the parameters of the channel model formulated in~\eqref{eq: channel-matrix}. It also changes the number of antenna elements that we can manufacture on the same antenna size, changing the antenna pattern and thereby the interference terms.
\end{remark}

\begin{remark}[Antenna Pattern]
Changing the antenna array affects only $\vec{a}_{\mathrm{UE}}$ and $\vec{a}_{\mathrm{BS}}$ in~\eqref{eq: channel-matrix}, and thereby the optimal precoding and combining vectors in~\eqref{eq: AnalogBeamforming}.
\end{remark}

\begin{remark}[Complexity of $\mathcal{P}_1$]\label{remark: complexity-of-full-coordination}
To optimally solve $\mathcal{P}_1$, we need CSI toward all UEs at all BSs of all operators. Given such knowledge, full intra- and inter-operator coordination is required to solve $\mathcal{P}_1$. In particular, all BSs should exchange their loads, current association solutions, and objective values.
\end{remark}

According to Remark~\ref{remark: complexity-of-full-coordination}, the BSs of every operator should be able to send/receive some pilot signals of all UEs of all operators, and exchange a huge amount of information with a central controller, which should then solve $\mathcal{P}_1$.
The complexity and cost of such level of channel estimation and coordination grow large with the number of BSs and UEs, and is in general overwhelming for mmWave networks with dense BS deployment. Moreover, if BSs or UEs belong to different network operators, a huge inter-operator signaling via the core networks is required for synchronization and for the calculation of $I_3$, see also~\cite{Naghshin2016Coverage,Pezeshki2016Anywhere}. We address these issues in the next subsection. Nonetheless, the solution of $\mathcal{P}_1$ gives a theoretical upper bound for the performance gain of spectrum sharing with analog precoders, which we further use as a benchmark for the performance of more practical sharing scenarios.



\subsection{Optimal Association with Sharing and No Inter-operator Coordination}\label{sec: optimal-association-semi-coordination-analog}
Inter-operator coordination is the main source of complexity of $\mathcal{P}_1$, which manifests itself in both calculating $I_3$ and exchanging the values of the objective functions.
To alleviate the cost and complexity of $\mathcal{P}_1$, we enforce the following design constraints. First, each operator maximizes only its own benefit without taking the objective of other operators into account. Second, the inter-cell interference $I_3$ is approximated by a quantity $\widehat{I}_3$ that depends only on $\vec{X}_z$ for each operator $z$. Consequently, $f_z$ can be calculated without any inter-operator coordination.
By substituting $\widehat{I}_3$ into~\eqref{eq: average-rate}, and the resulting $r_{b \hskip 0.2mm u}$ into~\eqref{eq: ObjectiveFunction1}, we pose the following optimization problem for each operator $z$:
\begin{subequations}\label{eq: long-term-resource-analog-p2}
\begin{alignat}{3}
\mathcal{P}_2:~& \underset{\vec{X}_z}{\text{maximize}} \hspace{3mm} && f_z (\vec{X}_z) \:,
\\
& {\text{subject to}} && \mbox{beamforming design~\eqref{eq: AnalogBeamforming}}\:,\\
\label{eq: const-unique-association-p2}
& && \sum\limits_{b \in \mathcal{B}_z} x_{b \hskip 0.2mm u} = 1 \:, \quad \forall u \in \mathcal{U}_z\:,
\\
\label{eq: cell-size-p2}
& && \sum\limits_{u \in \mathcal{U}_z} x_{b \hskip 0.2mm u} \leq N_{r} \:, \hspace{1.5mm} \forall b \in \mathcal{B}_z \:, \\
\label{eq: binary-variables-p2}
& && x_{b \hskip 0.2mm u} \in \{0,1 \} \:, \quad   \forall u \in \mathcal{U}_z, b \in \mathcal{B}_z \:,
\end{alignat}
\end{subequations}
which can be independently solved by individual operators. In the following, and without loss of generality, we assume $\widehat{I}_3 = 0$.

\begin{remark}[Complexity of $\mathcal{P}_2$]\label{remark: requiredInfoP2}
Optimization problem $\mathcal{P}_2$ requires neither inter-operator CSI knowledge nor inter-operator coordination. However, it still needs full CSI knowledge and coordination within one operator. Moreover, $\mathcal{P}_2$ can be solved by individual operators in parallel, and has substantially fewer dimensions than $\mathcal{P}_1$, thus its computational complexity is significantly less than that of $\mathcal{P}_1$.
\end{remark}

\begin{remark}[Optimality of $\mathcal{P}_2$]\label{remark: optimality-of-neglecting-I3}
If $I_3$ is small compared to the noise power, which occurs under large antenna regimes or sparse mmWave networks, the solution would not be too sensitive to the estimation error, namely ${|I_3|}$. However, as $I_3$ increases, deriving the optimal solution of $\mathcal{P}_1$ becomes more sensitive to the approximation error, making inter-network cooperation more important in order to avoid heavily suboptimal performance.
\end{remark}

Remark~\ref{remark: optimality-of-neglecting-I3} implies that interference-limited networks require a very good estimation of $I_3$, and without it, their performance can be severely degraded after spectrum sharing. This indeed reduces the benefits of spectrum sharing in traditional ``interference-limited'' cellular networks, as we will observe in Section~\ref{sec: assymptotic-performance}.
References~\cite{di2014stochastic,Shokri2015Transitional} have analyzed the interference level in mmWave networks.

\subsection{Practical Association with Sharing and No Coordination}
Many practical systems consider received signal strength indicator (RSSI)-based association, which requires neither inter- nor intra-operator coordination for the association phase. Besides this lack of coordination, RSSI-based association cannot balance the network load, which may lead to a substantial drop in both per-user throughput and fairness~\cite{andrews2014overview}. In short, each UE will be associated to the BS that can provide the highest received power. For benchmarking purposes, we will also consider this spectrum sharing option in the numerical results.

\subsection{Optimal Association with No Sharing}
In the no spectrum sharing scenario, we assume that the bandwidth $W$ is equally divided among all $Z$ operators. With universal frequency reuse, the bandwidth available to each BS is then $W_z = W/Z$. Due to non-overlapping bandwidths, the inter-operator interference is $I_3=0$. Then, similar to the previous subsection, for each operator $z$, $f_z$ depends only on $\vec{X}_z$. By substituting $W_z$ and $I_3=0$ into~\eqref{eq: average-rate}, and the resulting $r_{b \hskip 0.2mm u}$ into~\eqref{eq: ObjectiveFunction1}, we can formulate an association optimization problem for every operator $z$. This optimization problem, hereafter called $\mathcal{P}_3$, is very similar to~\eqref{eq: long-term-resource-analog-p2}, the only difference being the calculation of the objective function. Remarks~\ref{remark: requiredInfoP2} and~\ref{remark: optimality-of-neglecting-I3} hold for $\mathcal{P}_3$ as well. In particular, as $I_3 = 0$, inter-operator coordination will bring no gain to spectrum sharing.\footnote{Actually, the allocation of disjoint bands to different operators can be seen as a form of a priori and static inter-operator coordination.}

So far, we derived optimization problems that can characterize lower bounds for the performance gain of spectrum sharing, as their solutions are feasible. On the other hand, these solutions are suboptimal in general, since (i) analog precoders are used, which may not effectively cancel the multiuser interference and lack multiplexing gain, and (ii) optimal short-term time-frequency scheduling (which would further decrease the interference) is not applied. In order to provide an upper bound for the sharing performance, in the next section we consider digital precoding at the BS.


\section{Spectrum Sharing with Digital Beamforming}\label{sec: pooling-digital-beamforming}
In this section, we characterize the gain of spectrum sharing with digital precoding at the BSs when considering different coordination scenarios.
\subsection{SINR and Rate Models}
A digital precoder at the BSs enables MU-MIMO gain, namely a BS can serve all its UEs in the same time and frequency resources, as UEs are separable in the signal domain. The intra-cell interference to UE $u \in \mathcal{U}_z$ when being served by BS $b \in \mathcal{B}_z$ is
\begin{equation}\label{eq: intra-cell-interference-digital-precoder}
I_1 = \lambda_{b}\sum\limits_{j\in\mathcal{A}_{b}\setminus\{u\}}\left| \left(\vec{w}_{b \hskip 0.2mm u}^{\mathrm{UE}}\right)^{*} \vec{H}_{b \hskip 0.2mm u} \vec{w}_{bj}^{\mathrm{BS}} \right|^2 \:.
\end{equation}

Assuming that all the BSs are always concurrently serving all their own UEs (with MU-MIMO), we have that
\begin{equation}\label{eq: inter-cell-interference-digital-precoder}
I_2 = \sum\limits_{i \in \mathcal{B}_z \setminus \{b\}} \lambda_{b}\sum\limits_{j\in\mathcal{A}_{i}}\left| \left(\vec{w}_{b \hskip 0.2mm u}^{\mathrm{UE}}\right)^{*} \vec{H}_{iu} \vec{w}_{ij}^{\mathrm{BS}} \right|^2 \:.
\end{equation}
Clearly, inter-cell coordination can affect the choice of set $\mathcal{A}_{i}$ and thereby $I_2$. However, as we will show in Section~\ref{sec: numerical-results}, $I_2$ is very small in general for sufficiently large $N_{\mathrm{BS}}$ and $N_{\mathrm{UE}}$. This interesting property reduces the importance and necessity of inter-cell coordination in large-antenna regimes.

The inter-operator interference is
\begin{equation}\label{eq: inter-operator-interference-digital-precoder}
I_3 = \sum\limits_{k=1 \hfill \atop k \neq z \hfill}^{Z} \, \sum\limits_{i \in \mathcal{B}_k} \lambda_{i}\sum\limits_{j\in\mathcal{A}_{i}}\left| \left(\vec{w}_{b \hskip 0.2mm u}^{\mathrm{UE}}\right)^{*} \vec{H}_{iu} \vec{w}_{i j}^{\mathrm{BS}} \right|^2 \:.
\end{equation}
Then, the average rate that UE $u$ can get from BS $b$ can be computed from~\eqref{eq: average-rate}, and the long-term rate of UE $u$ is again $r_{u} = \sum_{b \in \mathcal{B}_z}{x_{b \hskip 0.2mm u}r_{b \hskip 0.2mm u}}$.

\subsection{Optimal Association with Sharing and Full Coordination}\label{sec: optimal-association-full-coordination-digital}
Suppose full CSI knowledge in the entire network and full information sharing (e.g., all BSs know the analog combiners of all the UEs). Define the following effective channel matrix (processed by an analog combiner) between BS $b$ and UE $u$
\begin{equation}\label{eq: effective-channel}
\overline{\vec{H}}_{b u} = \begin{bmatrix}
\vdots \\
\vspace{-2mm} \\
\left(\vec{w}_{b \hskip 0.2mm u}^{\mathrm{UE}}\right)^{*} \vec{H}_{iu}\\
\vspace{-2mm}\\
\vdots
\end{bmatrix} \in \mathds{C}^{|\mathcal{B}|\times N_{\mathrm{BS}}},  \hspace{2mm} \hspace{-2mm}\begin{array}{ll}
\forall k \in\{1,\ldots, Z\},\\ \forall  i \in \mathcal{B}_k \:.
\end{array}
\end{equation}
Then, define
\begin{equation}\label{eq: effective-channel2}
\overline{\vec{H}}_{z} =
\begin{bmatrix}
\overline{\vec{H}}_{1 1}^{*} & \hspace{-1.1mm} \cdots & \hspace{-1.1mm} \overline{\vec{H}}_{1 |\mathcal{A}_{1}|}^{*} &
\hspace{-1.1mm} \overline{\vec{H}}_{2 1}^{*} & \hspace{-1.1mm} \cdots &\hspace{-1.1mm} \overline{\vec{H}}_{2 |\mathcal{A}_{2}|}^{*} & \hspace{-1.1mm}
\cdots & \hspace{-1.1mm} \overline{\vec{H}}_{bu}^{*} & \hspace{-1.1mm}\cdots
\end{bmatrix}^{*} \hspace{-1mm} ,
\end{equation}
for all $b \in \mathcal{B}_z$ and $u \in \mathcal{A}_{b}$. Finally, define the following effective channel matrix ${\overline{\vec{H}} \in \mathds{C}^{|\mathcal{U}||\mathcal{B}|\times N_{\mathrm{BS}}}}$:
\begin{equation}\label{eq: effective-channel3}
\overline{\vec{H}} = \begin{bmatrix}
\overline{\vec{H}}_{1} \\ \overline{\vec{H}}_{2} \\ \vdots \\
\overline{\vec{H}}_{Z}
\end{bmatrix} \:.
\end{equation}
Suppose that UE $u$ is being served by BS $b$, and that $\vec{H}_{b \hskip 0.2mm u}$ has appeared in row $m$ of $\overline{\vec{H}}$. In MRT, the precoding vector of UE $u$ is the transpose conjugate of row $m$ of $\overline{\vec{H}}$. In RZF, the precoding vector of UE $u$ is column $m$ of $(\overline{\vec{H}} + c\vec{I})^{\dag}$. The analog combiners at the UEs are then calculated using~\eqref{eq: DigitalBeamforming}.

As in Section~\ref{sec: optimal-association-full-coordination-analog}, we consider a logarithmic utility function $f_z$ for each operator $z$ to ensure both high network throughput and fairness among its individual UEs, namely $f_z (\vec{X}) = \sum_{u \in \mathcal{U}_z}\,{\log \left(\sum_{b \in \mathcal{B}_z} \, x_{b \hskip 0.2mm u}r_{b \hskip 0.2mm u}\right)}$. To find the optimal association, we can formulate the following multi-objective optimization problem:
\begin{subequations}\label{eq: long-term-resource-digital}
\begin{alignat}{3}
\mathcal{P}_4:~& \underset{\vec{X}}{\text{maximize}} \hspace{3mm} && \left[f_1 (\vec{X}), f_2 (\vec{X}), \ldots,f_{Z} (\vec{X})\right] \:, \\
& {\text{subject to}} && \mbox{beamforming design~\eqref{eq: DigitalBeamforming}}\:,\\
\label{eq: const-unique-association-binary-variables}
& && \mbox{\eqref{eq: const-unique-association}, \eqref{eq: cell-size-p1}, \eqref{eq: binary-variables}, and~\eqref{eq: binary-variables-11}} \:,
\end{alignat}
\end{subequations}
where $N_r = N_{\mathrm{BS}}$ in constraint~\eqref{eq: cell-size-p1}. The solution of $\mathcal{P}_4$ regulates long-term load balancing of the network.

\begin{remark}[Complexity of $\mathcal{P}_4$]\label{remark: complexity-of-full-coordination-DBF}
To optimally solve $\mathcal{P}_4$, we need full coordination among all operators as well as knowledge of CSI toward all UEs at all BSs.
\end{remark}

Remark~\ref{remark: complexity-of-full-coordination-DBF} indicates that finding the optimal solution of $\mathcal{P}_4$ may require formidable complexity and cost for channel estimation and for coordination. Nonetheless, it gives a benchmark on the performance of spectrum sharing schemes with digital precoders.
In particular in the case of RZF precoding, it is easy to show that $I_1$, $I_2$, and $I_3$ can be made arbitrarily close to zero in the large antenna regime by properly choosing the regularization term $c$ in Equation~\eqref{eq: digital-beamforming-C2}, as suggested by~\cite{wagner2012large}.

\subsection{Optimal Association with Sharing and No Inter-operator Coordination}
To mitigate the necessity of inter-operator CSI knowledge and coordination, we apply similar modifications to $\mathcal{P}_4$ as in Section~\ref{sec: optimal-association-semi-coordination-analog}. Let $\widehat{I}_3$ denote an approximation to the inter-cell interference $I_3$. Using CSI knowledge and information exchange within one operator, each BS $b$ calculates
\begin{equation}\label{eq: effective-channel4}
\overline{\vec{H}}_{b  u} = \begin{bmatrix}
\vdots \\
\vspace{-2mm} \\
\left(\vec{w}_{b \hskip 0.2mm u}^{\mathrm{UE}}\right)^{*} \vec{H}_{iu}\\
\vspace{-2mm}\\
\vdots
\end{bmatrix} \in \mathds{C}^{|\mathcal{B}_z|\times N_{\mathrm{BS}}},
\forall i \in \mathcal{B}_z \:,
\end{equation}
and each operator $z$ independently calculates its own effective channel $\overline{\vec{H}}_z$ using~\eqref{eq: effective-channel2}. Then, it computes the digital precoder vectors similar to the previous subsection assuming $\overline{\vec{H}} = \overline{\vec{H}}_z$. The analog combiners are then calculated using~\eqref{eq: DigitalBeamforming}. Therefore, the beamforming design of each operator becomes independent of the others. The price, however, is a higher inter-operator interference, compared to the full information sharing scenario. We pose the following optimization problem for each operator $z$:
\begin{subequations}\label{eq: long-term-resource-digital-2}
\begin{alignat}{3}
\mathcal{P}_5:~ &\underset{\vec{X}_z}{\text{maximize}} \hspace{3mm} &&f_z (\vec{X}_z) \:, \\
& {\text{subject to}} && \mbox{beamforming design~\eqref{eq: DigitalBeamforming}}\:,\\
\label{eq: cell-size-p5}
& &&\sum\limits_{b \in \mathcal{B}_z} x_{b \hskip 0.2mm u} \leq N_{\mathrm{BS}} \:, \quad \forall b \in \mathcal{B}_z\:, \\
& && \mbox{ \eqref{eq: const-unique-association-p2}, \eqref{eq: cell-size-p2}, and \eqref{eq: binary-variables-p2}} \:.
\end{alignat}
\end{subequations}
which can be independently solved by each individual operator in parallel. Note that Remarks~\ref{remark: requiredInfoP2} and~\ref{remark: optimality-of-neglecting-I3} on the complexity and optimality of $\mathcal{P}_2$ still hold here. Namely, optimization problem $\mathcal{P}_5$ requires intra-operator but no inter-operator coordination; however, the performance of spectrum sharing among multiple interference-limited networks (i.e., traditional cellular networks) may be severely reduced without inter-operator coordination.

\subsection{Practical Association with Spectrum Sharing and No Coordination}
For this case, we assume that all operators are using the whole bandwidth, so $W_z = W$. Within every operator $z$, UE $u$ will be associated to BS $b$ that can provide the highest received power.
As each BS $b$ has only CSI knowledge of its own UEs, it computes the following effective channel ${\overline{\vec{H}}_{b u} = \left(\vec{w}_{b \hskip 0.2mm u}^{\mathrm{UE}}\right)^{*} \vec{H}_{b \hskip 0.2mm u}}$, and then calculates its own effective channel $\overline{\vec{H}}_{b}$ using~\eqref{eq: effective-channel2}. Then, it assumes $\overline{\vec{H}} = \overline{\vec{H}}_{z} = \overline{\vec{H}}_{b}$ and computes the digital precoder vectors similar to the previous subsections. The analog combiners are calculated using~\eqref{eq: DigitalBeamforming}.

\subsection{Optimal Association with No Spectrum Sharing}
Without bandwidth sharing, each BS has $W_z = W/Z$ bandwidth, and the inter-operator interference is $I_3=0$. The effective channel is identical to~\eqref{eq: effective-channel4}. By substituting $W_z$ and $I_3$ into~\eqref{eq: average-rate}, and the resulting $f_z (\vec{X}_z)$ into~\eqref{eq: long-term-resource-digital-2}, we can formulate an association optimization problem for every operator $z$. This optimization problem, hereafter called $\mathcal{P}_6$, is very similar to~\eqref{eq: long-term-resource-digital-2}, the only difference being the calculation of the objective function. Remark~\ref{remark: requiredInfoP2} holds for $\mathcal{P}_6$ as well.

The solutions of optimization problems $\mathcal{P}_1$--$\mathcal{P}_6$ characterize the gain of coordination and beamforming for the performance of spectrum sharing schemes in mmWave networks. In particular, comparing the solutions of $\mathcal{P}_1$ and $\mathcal{P}_2$ (or equivalently $\mathcal{P}_4$ and $\mathcal{P}_5$) shows the gain of inter-operator coordination with an analog (or digital) precoding scheme. Similarly, the solutions of $\mathcal{P}_2$ and $\mathcal{P}_3$ (or equivalently $\mathcal{P}_5$ and $\mathcal{P}_6$) enable quantifying the benefits of intra-operator coordination for spectrum sharing.

In Section~\ref{sec: numerical-results}, we will provide numerical evaluations based on the optimization problem formulations given here, which will allow us to draw important design insights and conclusions.

\section{Asymptotic Performance Analysis}\label{sec: assymptotic-performance}
In this section, we evaluate the asymptotic behavior of spectrum sharing with analog and digital precoders for a large number of antennas. We start with the following lemma:
\begin{lemma}\label{lemma: asymptotic-zero-product-aBS}
Consider the antenna response in Equation~\eqref{eq: ULA-antenna-response}. Then, $\limsup_{N_{\mathrm{BS}} \to \infty } \vec{a}_{\mathrm{BS}}^{*}(\theta) \: \vec{a}_{\mathrm{BS}}(\phi) = 1$ if $\theta = \phi$, and 0 otherwise. Similarly, $\limsup_{N_{\mathrm{UE}} \to \infty }\vec{a}_{\mathrm{UE}}^{*}(\theta) \: \vec{a}_{\mathrm{UE}}(\phi) = 1$ if $\theta = \phi$, and 0 otherwise. Moreover, if BSs, UEs, and scatterers are randomly independently located in the environment, $\theta \neq \phi$ almost surely.
\end{lemma}
\begin{IEEEproof}
Using the definition of antenna response given in Equation~\eqref{eq: ULA-antenna-response}, $\vec{a}_{\mathrm{BS}}^{*}(\theta)$ and $\vec{a}_{\mathrm{BS}}^{*}(\phi)$ form different columns of a discrete Fourier transform (DFT) matrix as $N_{\mathrm{BS}} \to \infty$. Noting that the DFT matrix is a unitary matrix, the first part of the lemma follows. For the second part, random and independent locations of the objects (BSs, UEs, and scatterers) imply that $\theta$ and $\phi$ are i.i.d. absolutely continuous random variables. Therefore, $\theta \neq \phi$ almost surely, which concludes the proof.
\end{IEEEproof}
Lemma~\ref{lemma: asymptotic-zero-product-aBS} implies that vector $\vec{a}_{\mathrm{BS}}(\theta)$ (or $\vec{a}_{\mathrm{UE}}(\theta)$) with different $\theta$ creates an asymptotically orthonormal basis, which can be used as orthogonal spatial signatures. More interestingly, asymptotically, there are infinitely many such spatial signatures (realized by changing $\theta$) almost surely, and there will be no multiuser interference if one gives different signatures to different UEs (BSs). We use this important property in the next subsections.

\subsection{Analog Precoding}\label{sec: asymptotic-results-ABF}
Using Lemma~\ref{lemma: asymptotic-zero-product-aBS}, it is easy to show the following asymptotic result for the solution of~\eqref{eq: AnalogBeamforming}:

\begin{lemma}[\hspace{-0.1mm}{\cite[Corollary~4]{Ayach2012Capacity}}]\label{lemma: asymptotic-solution-for-ABF}
Consider the mmWave channel model in Equation~\eqref{eq: channel-matrix}. In the limit of large $N_{\mathrm{BS}}$ and infinite resolution for the BS precoding codebook $\mathcal{W}^{\mathrm{BS}}$, the solution of optimization problem~\eqref{eq: AnalogBeamforming} is $\vec{w}_{b \hskip 0.2mm u}^{\mathrm{BS}} = \vec{a}_{\mathrm{BS}}\left(\theta_{b \hskip 0.2mm u \hskip 0.2mm n^{*}}^{\mathrm{BS}}\right)$, where $\theta_{b \hskip 0.2mm u \hskip 0.2mm n^{*}}^{\mathrm{BS}}$ is the AoD of path $n^{*} = \arg\max_n |g_{b \hskip 0.2mm u \hskip 0.2mm n}|$. Similarly, for large $N_{\mathrm{UE}}$ and infinite resolution for the UE combiner codebook $\mathcal{W}^{\mathrm{UE}}$, the solution of optimization problem~\eqref{eq: AnalogBeamforming} is $\vec{w}_{b \hskip 0.2mm u}^{\mathrm{UE}} = \vec{a}_{\mathrm{UE}}\left(\theta_{b \hskip 0.2mm u \hskip 0.2mm n^{*}}^{\mathrm{UE}}\right)$, where $\theta_{b \hskip 0.2mm u \hskip 0.2mm n^{*}}^{\mathrm{UE}}$ is the AoA of path $n^{*}$.
\end{lemma}
Lemma~\ref{lemma: asymptotic-solution-for-ABF} implies that, in the asymptotic regime, the optimal precoding and combining vectors are spatial signatures toward the strongest path between UE $u$ and BS $b$. Since the channel gain may change per coherence interval, the precoding vector may change in every coherence time. However, instead of being any arbitrary vector from $\mathcal{W}^{\mathrm{BS}}$, it can be chosen among only $N_{b \hskip 0.2mm u}$ vectors (AoDs of the channels). Similarly, the combining vector at the UE can be chosen among only $N_{b \hskip 0.2mm u}$ vectors in $\mathcal{W}^{\mathrm{UE}}$.

\begin{remark}\label{remark: Single-path-ABF}
Consider a single path between any BS-UE pairs, namely $N_{b \hskip 0.2mm u} = 1$ for all $b$ and $u$. In the limit of large $N_{\mathrm{BS}}$ and infinite resolution for BS precoding codebook $\mathcal{W}^{\mathrm{BS}}$, the optimal analog precoder depends only on the second order statistics of the channel, namely AoDs.
Similarly, for large large $N_{\mathrm{UE}}$ and infinite resolution for UE combiner codebook $\mathcal{W}^{\mathrm{UE}}$, the optimal analog combiner depend only on the AoAs.
\end{remark}

Noting that the second order statistics typically remain unchanged for many coherence times, Remark~\ref{remark: Single-path-ABF} describes the conditions under which analog beamforming and association can take place at similar time scales. These conditions most probably hold in mmWave networks where the wireless channel is sparse in the angular domain, and the number of antenna elements is high enough due to the small wavelength~\cite{Rappaport2014mmWaveBook}. Therefore, Remark~\ref{remark: Single-path-ABF} has the following two implications: first, it enables the possibility of low complexity design of analog precoder and combiner (once per many coherence times), as also suggested in~\cite{Nam2014Joint,Bogale2015Hybrid}; second, it enables the possibility of designing analog beamforming as a part of long-term resource allocation policies, as also suggested in~\cite{shokri2015mmWavecellular}. We will further elaborate on Remark~\ref{remark: Single-path-ABF} in Section~\ref{sec: numerical-results}. In the asymptotic case, we also have the following remark:

\begin{remark}\label{remark: asymptotic-no-int}
Suppose that perfect CSI is available. The interference components, formulated in Equations~\eqref{eq: intra-cell-interference}--\eqref{eq: inter-operator-interference}, vanish almost surely as either $N_{\mathrm{BS}} \to \infty$ or $N_{\mathrm{UE}} \to \infty$.
\end{remark}
In general, intra- and inter-operator coordinations reduce different interference components and also balance the network load. Remark~\ref{remark: asymptotic-no-int} suggests that intra- and inter-operator coordinations brings no additional gain to the interference reduction in the large antenna regime. Zero inter-operator interference implies that $f_z(\vec{X}) = f_z(\vec{X}_z)$, and therefore optimization problem $\mathcal{P}_1$ can be decomposed into $Z$ independent optimization problems, one for each operator, with no penalty on the performance. In other words, $\mathcal{P}_1$ and $\mathcal{P}_2$ give the same optimal association solution, which leads to the following conclusion: \emph{if either $N_{\mathrm{BS}} \to \infty$ or $N_{\mathrm{UE}} \to \infty$, the performance of a system with full inter-operator coordination becomes identical to that of a system with no such coordination.}
Nonetheless, intra-operator coordination can be still beneficial due to the load balancing gain.\footnote{With the conventional RSSI-based BS association rule (with no load balancing gain) and in the large antenna regime, the performance of an uncoordinated (no intra-operator coordination) system tends to that of the coordinated system, due to zero intra-operator interference, as also highlighted in the seminal work by Marzetta~\cite{marzetta2010noncooperative}.}

\subsection{Digital Precoding}
As the UEs are still using an analog combiner, all the Lemmas and Remarks of the previous subsection hold for the UE, namely they are valid as $N_{\mathrm{UE}} \to \infty$. In particular, we have the following remark:
\begin{remark}\label{remark: Asymptotic-digital1}
Consider the mmWave channel model in Equation~\eqref{eq: channel-matrix}. In the limit of large $N_{\mathrm{UE}}$ and infinite resolution for combining codebook, $\vec{w}_{b \hskip 0.2mm u}^{\mathrm{UE}} = \vec{a}_{\mathrm{UE}}\left(\theta_{b  \hskip 0.3mm  u  \hskip 0.3mm n^{*}}^{\mathrm{BS}}\right)$, where $n^{*} = \arg\max_n |g_{b \hskip 0.3mm u  \hskip 0.3mm n}|$.
\end{remark}
The proof of this remark is very similar to the proof of~\cite[Corollary~4]{Ayach2012Capacity}, which is essentially based on the asymptotic orthogonality of the spatial signatures (Lemma~\ref{lemma: asymptotic-zero-product-aBS}), and therefore it is omitted. Remark~\ref{remark: Asymptotic-digital1} implies the following important property:
\begin{prop}\label{prop: Asymptotic-digital2}
Suppose $N_{b u} = 1$ for all $b \in \mathcal{B}$ and $u \in \mathcal{U}$. In the limit of large $N_{\mathrm{UE}}$, large $N_{\mathrm{BS}}$, and infinite resolution for combining codebook, both MRT and RZF precoders provide the same rate, which is equal to the maximum achievable rate, obtained by dirty paper coding. This is valid for all information sharing scenarios: full coordination, only intra-operator coordination, and no coordination scenarios.
\end{prop}


\begin{IEEEproof}
If the channels among one BS and all other UEs in the entire network become orthogonal, then inverting the channel matrix (RZF) becomes a rotation operation (MRT). This implies the same asymptotic performance for MRT and RZF. Moreover, there is no loss in the channel inversion in this case. So, the achievable rate of BS $b$ (with both MRT and RZF) can linearly increase with $\mathcal{A}_b$, which is as promised by dirty paper coding. To complete the proof, we only need to show orthogonality of the channels as $N_{\mathrm{BS}} \to \infty$. If we represent channel model~\eqref{eq: channel-matrix} in the matrix form
\begin{equation}
\vec{H}_{b \hskip 0.2mm u} = \vec{A}_{b \hskip 0.2mm u}^{\mathrm{UE}} \, \vec{G}_{b \hskip 0.2mm u} \, \left(\vec{A}_{b \hskip 0.2mm u}^{\mathrm{BS}} \right)^* \:,
\end{equation}
where $\vec{A}_{b \hskip 0.2mm u}^{\mathrm{UE}}$ and $\vec{A}_{b \hskip 0.2mm u}^{\mathrm{BS}}$ contain the array response vectors toward all the AoAs and AoDs, and $\vec{G}_{b \hskip 0.2mm u}$ is a diagonal matrix with the corresponding path gains, then we have
\begin{equation}
\vec{H}_{b \hskip 0.2mm u} \vec{H}_{ij}^{*} = \vec{A}_{b \hskip 0.2mm u}^{\mathrm{UE}} \, \vec{G}_{b \hskip 0.2mm u} \, \left(\vec{A}_{b \hskip 0.2mm u}^{\mathrm{BS}} \right)^* \, \vec{A}_{ij}^{\mathrm{BS}} \, \vec{G}_{ij}^* \, \left(\vec{A}_{ij}^{\mathrm{UE}} \right)^* \:,
\end{equation}
where $\left(\vec{A}_{b \hskip 0.2mm u}^{\mathrm{BS}} \right)^* \vec{A}_{ij}^{\mathrm{BS}}$ is almost surely a zero matrix if ${N_{\mathrm{BS}} \to \infty}$; see Lemma~\ref{lemma: asymptotic-zero-product-aBS}. This implies almost sure orthogonality of the channels, which completes the proof.
\end{IEEEproof}

As illustrated in its proof, the main reason for Remark~\ref{remark: Asymptotic-digital1} is the orthogonality of spatial signatures of individual BS-UE pairs. This orthogonality appears also in massive MIMO scenarios~\cite{hoydis2013massive} and in massive UEs scenarios~\cite{yoo2006optimality}, where we can always find a subset of UEs (among infinitely many of them) with orthogonal channels. In all these cases, both MRT and RZF are optimal.

\begin{remark}\label{remark: SameSolution}
For an interference-free network, optimization problems~$\mathcal{P}_1$ and~$\mathcal{P}_2$ have the same solution. The same is true for $\mathcal{P}_4$ and $\mathcal{P}_5$.
\end{remark}


So far, we used ideal channel estimation, beamforming, and coordination schemes, with the aim of providing upper bounds for the performance of spectrum sharing. In the next subsection, we discuss the impact of real-world effects on the performance of spectrum sharing.

\subsection{Impact of Quantized Codebooks, Imperfect CSI, and Limited Feedback}
Our asymptotic analysis assumes precoding and combining codebooks ($\mathcal{W}^{\mathrm{BS}}$ and $\mathcal{W}^{\mathrm{UE}}$ respectively) of infinite resolution to find the precoding and combining vectors in closed-form and to obtain useful insights on the ultimate performance. In practice, due to the constraints of RF hardware, including the availability of quantized angles for RF phase shifters, only quantized beamforming codebooks are feasible. However, the simulation results in \cite{Alkhateeb:15} indicate that the precoding gain is not very sensitive to angular quantization. Therefore, we argue that for the purpose of analyzing the performance of spectrum sharing systems in the asymptotic regime, the infinite resolution codebook assumption does not introduce a significant error to the problem, and the ultimate insights are valid even for a codebook with finite resolution (assuming a sufficient resolution).

Also, recall that the problem of finding the precoding and combining vectors when using analog and fully digital precoding, formulated in \eqref{eq: AnalogBeamforming} and \eqref{eq: DigitalBeamforming},
requires perfect CSI knowledge. In practice, channel estimation in mmWave systems is difficult, because of the large channel dimensions and because the effective channel seen at baseband after analog filtering may have a lower rank than the actual channel, thereby not allowing full recovery of the actual channel. However, recent works exploiting the sparsity of the channel developed compressed sensing-based techniques that can be leveraged to acquire CSI both at the transmitter and at the receiver \cite{Mendez-Rial:15}.

As it is visible in the geometric channel model~\eqref{eq: channel-matrix},
estimating the mmWave channel implies estimating the parameters (AoA, AoD and random channel gain) of the $N_{b \hskip 0.2mm u}$ propagation paths between the BS and the UE.
To this end, several previous works proposed channel estimation algorithms applicable for single-path and sparse multi-path mmWave channels \cite{Alkhateeb:14}, \cite{Alkhateeb:15}.
Performance studies in \cite{Alkhateeb:15} indicate that hybrid analog/digital precoding algorithms for downlink multi-user mmWave systems that assume the availability of a limited feedback channel between the UE and the BS achieve good performance compared to the digital unconstrained precoding schemes. Therefore, although the precoding algorithms developed in our paper assume perfect channel knowledge, the obtained numerical results in the next section can be regarded as reasonable approximate upper bounds for the achievable rates.

Finally, in practice, a BS using hybrid analog/digital precoding and a large antenna array to serve multiple UEs needs a feedback channel to acquire CSI at the transmitters and at the receivers. Due to recent advances in limited feedback hybrid precoding systems, the capacity of the feedback channel is not expected to form a bottleneck in single path or sparse multipath mmWave systems \cite{Alkhateeb:14,Alkhateeb:15,Mendez-Rial:15}. The investigation of the inherent trade-off between the training and feedback overhead and the achievable performance in spectrum sharing systems is left for future studies.

\section{Numerical Results}\label{sec: numerical-results}
In this section, based on the analytical characterization provided in the previous sections, we numerically illustrate the feasibility of spectrum sharing in mmWave networks. We assume that BSs and UEs are randomly distributed on the plane according to independent Poisson point processes. Without loss of generality, we assume four operators, and a total bandwidth of 2~GHz. We consider three bandwidth sharing scenarios:
\begin{itemize}
  \item Exclusive: each operator uses a 500~MHz exclusive bandwidth;
  \item Partial bandwidth sharing: operators 1 and 2 share the first 1~GHz, and operators 3 and 4 share the second 1~GHz; and
  \item Full bandwidth sharing: all four operators share the whole 2~GHz bandwidth.
\end{itemize}
We recall that we use an ideal coordination scheme. In particular, we consider a centralized coordination approach where a central entity collects all the required information, solves the association optimization problem, and broadcasts the solution. Moreover, we assume that all BSs are synchronized, and that there is no delay in the interface between the BSs and the central entity. We also consider $N_r = 6$ RF chains at each BS and only one RF chain at each UE. Unless otherwise mentioned, we consider a 32~GHz carrier frequency, a BS density of 100~BSs/km$^2$ for every operator, a user density of 600 UEs/km$^2$, and 25~dBm total transmission power at each BS. For the spectrum sharing scenarios with no inter-operator coordination ($\mathcal{P}_2$ and $\mathcal{P}_4$), we consider $\widehat{I}_3 = 0$ and find the corresponding optimal association. However, note that the actual value of $I_3$ is considered when computing the performance of that association. We simulate 100 random topologies and find the optimal association and beamforming for every topology.

\subsection{Analog Precoding}
\begin{figure}[!t]
  \centering
  \begin{subfigure}[t]{\columnwidth}
    \centering
    \includegraphics[width=\columnwidth]{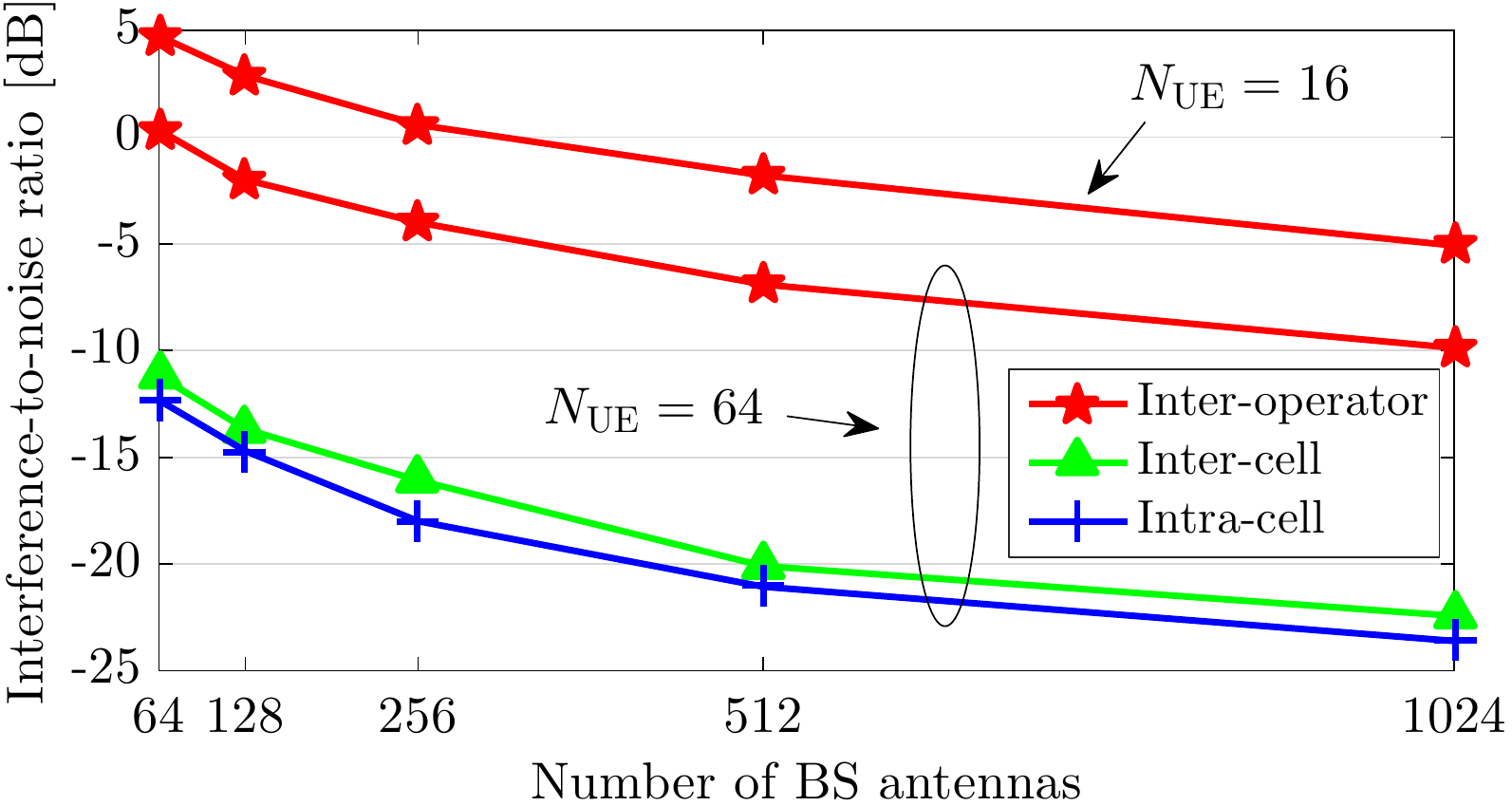}
    \caption{\label{subfig: Interference_average}}
  \end{subfigure}
  \begin{subfigure}[t]{\columnwidth}
       \centering
	\includegraphics[width=\columnwidth]{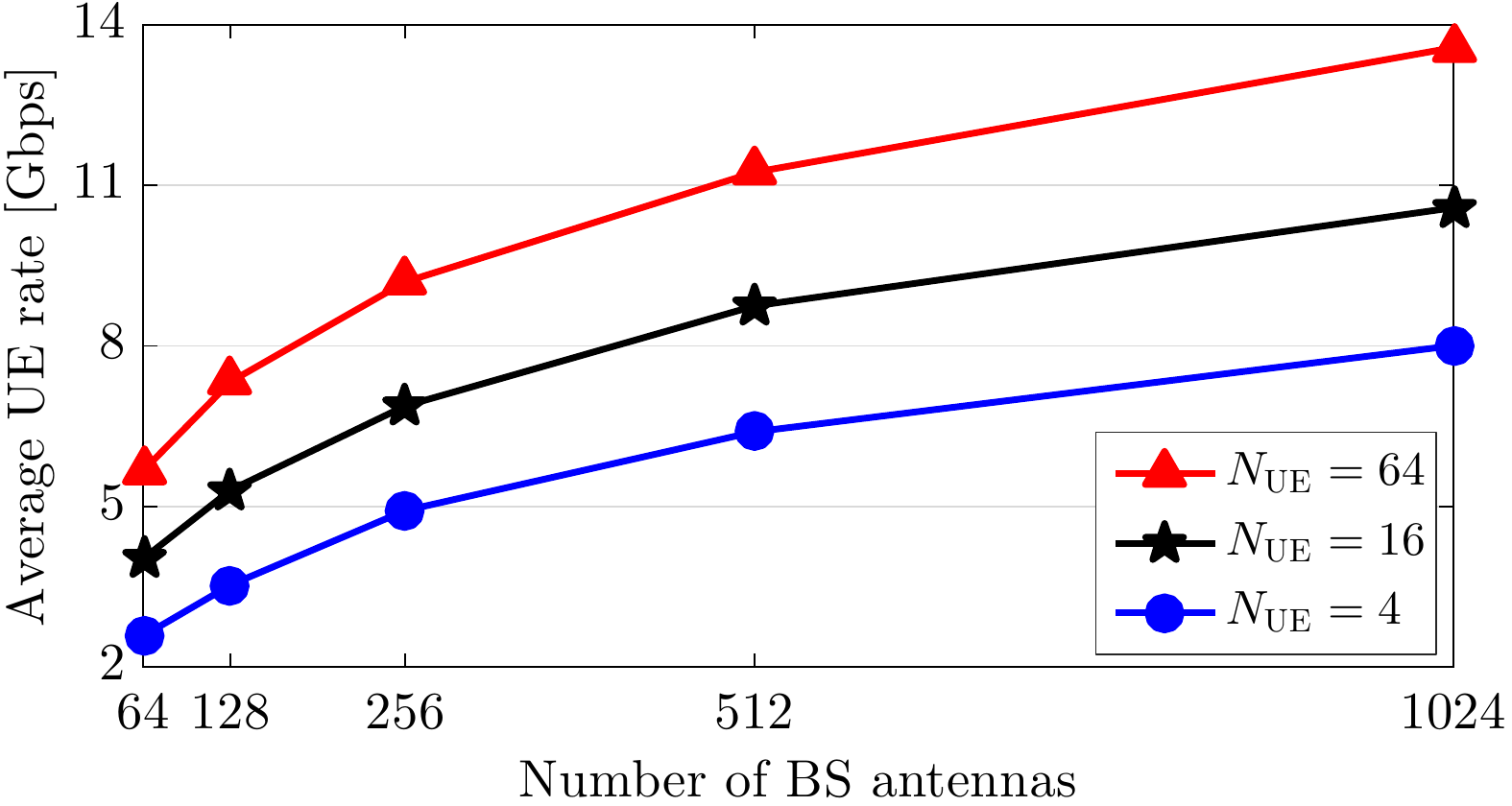}
    \caption{\label{subfig: Rate_average_P1}}
    \end{subfigure}

  \caption{Average interference and rate performance for full information and bandwidth sharing scenario with analog precoding, $\mathcal{P}_1$.}
  \label{fig: AveragePerformanceP1}
\end{figure}
We first study the performance of full information and bandwidth sharing scenario with analog precoding, i.e., optimization problem $\mathcal{P}_1$. Fig.~\ref{subfig: Interference_average} illustrates the average interference level, normalized to the noise power. Generally, inter-operator interference is the dominant term in the aggregated interference, as more BSs are contributing to this term.
All the interference components vanish when $N_{\mathrm{BS}}$ grows large, as predicted in Remark~\ref{remark: asymptotic-no-int}. Moreover, the number of antennas at the UEs has a complementary effect to the number of BS antennas, in terms of interference (i.e., the amount of interference roughly depends on the product of $N_{\mathrm{BS}}$ and $N_{\mathrm{UE}}$). In particular, this figure shows that a network with $(N_{\mathrm{BS}} = 256, N_{\mathrm{UE}} = 64)$ has almost the same average interference as a network with $(N_{\mathrm{BS}} = 1024, N_{\mathrm{UE}} = 16)$. This observation suggests that adding a few more antennas at the UEs (if their space and battery allow) may have the same effect as adding a large number of antennas at the BSs. Fig.~\ref{subfig: Rate_average_P1} shows the average rate of the UEs against the number of BS antennas.
Increasing either $N_{\mathrm{BS}}$ or $N_{\mathrm{UE}}$ improves the average rate due to both higher antenna gains and lower multiuser interference, but at a logarithmic speed. In other words, for high enough SINR, adding more antenna elements may bring negligible rate improvement. To further enhance the rate, we can either transmit parallel streams to one UE or add more bandwidth to the system.

\begin{figure}[t]
  \centering
  \includegraphics[width=\columnwidth]{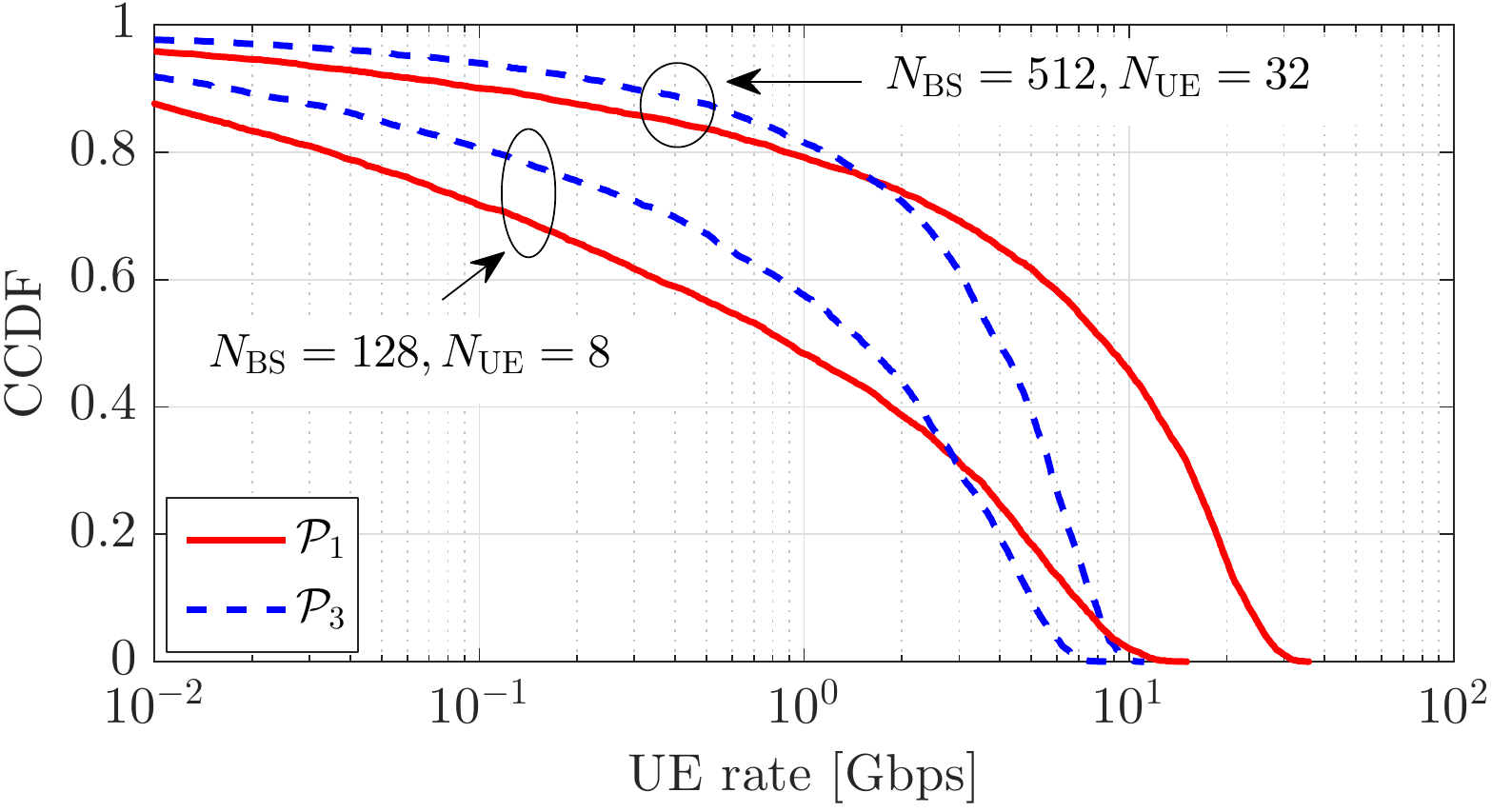}

  \caption{CCDF of the UE rate with analog precoding.}\label{fig: RateCCDF_P1_P3}
\end{figure}
Fig.~\ref{fig: RateCCDF_P1_P3} depicts the complementary cumulative distribution function (CCDF) of the UEs rates achieved by the association of $\mathcal{P}_1$ and $\mathcal{P}_3$ (no sharing). From this figure, the benefits of spectrum sharing are seen to heavily depend on the number of antenna elements both at the BSs and at the UEs. In the example of this paper, spectrum sharing is not beneficial with relatively small $N_{\mathrm{BS}}$ and $N_{\mathrm{UE}}$. However, with sufficiently large antenna arrays, spectrum sharing would be beneficial for the majority of the UEs, as can be seen from the $(N_{\mathrm{BS}} = 512, N_{\mathrm{UE}} = 32)$ curve. The main reason is that, though asymptotically able to cancel multiuser interference, pure analog beamforming does not have a good interference rejection performance in the non-asymptotic regime, as opposed to digital precoding that can almost completely reject the interference even in non-asymptotic regime.

\begin{figure}[!t]
  \centering
  \begin{subfigure}[t]{\columnwidth}
    \centering
    \includegraphics[width=\columnwidth]{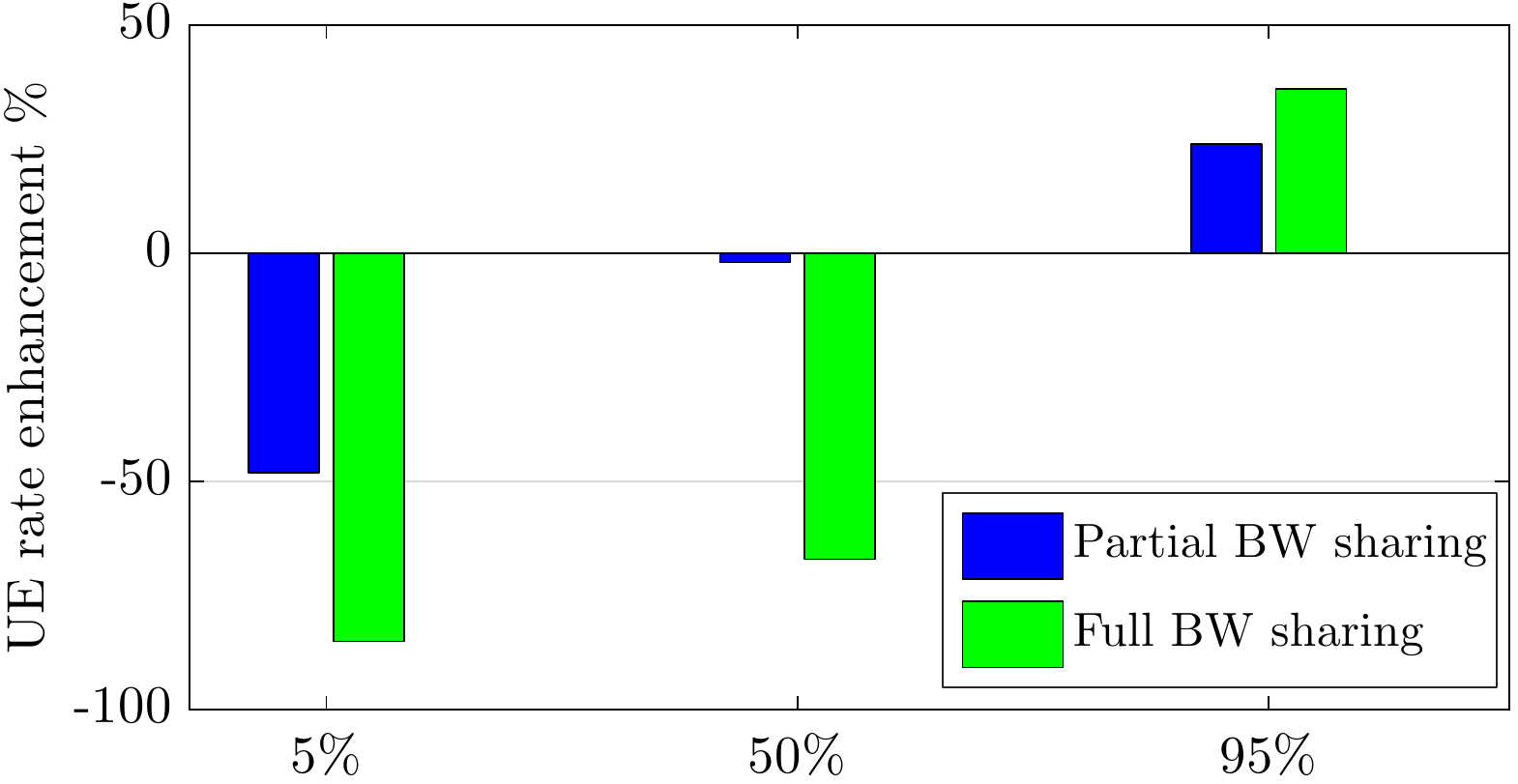}
    \caption{$N_{\mathrm{BS}} = 256, N_{\mathrm{UE}} = 2$}\label{subfig: Rate_percentile_P2_SpectrumSharing_BS128_UE1}
  \end{subfigure}
  \begin{subfigure}[t]{\columnwidth}
       \centering
	\includegraphics[width=\columnwidth]{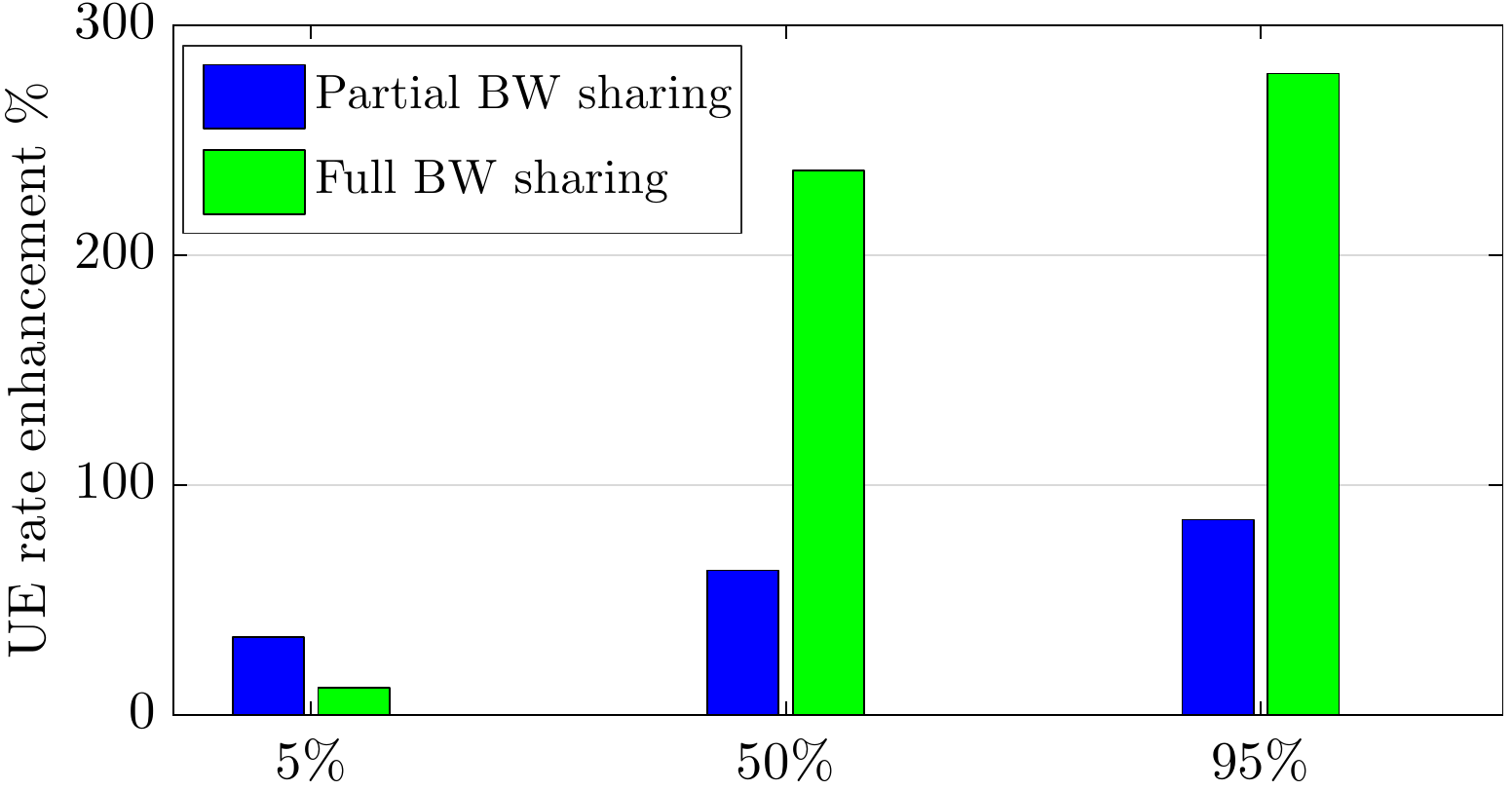}
    \caption{$N_{\mathrm{BS}} = 1024, N_{\mathrm{UE}} = 64$}\label{subfig: Rate_percentile_P2_SpectrumSharing_BS1024_UE128}
    \end{subfigure}

  \caption{Bandwidth (BW) sharing performance at 32~GHz, under the assumption of no inter-operator coordination and analog precoding. The baseline is a system with exclusive spectrum allocation.}
  \label{fig: Rate_percentile_P2}
\end{figure}
Fig.~\ref{fig: Rate_percentile_P2} analyzes the contributions of bandwidth sharing and of massive antennas on the performance of spectrum sharing. We show the 5th, 50th, and 95th percentiles of the UE downlink rates. The baseline is a system with exclusive spectrum allocation. We observe in Fig.~\ref{subfig: Rate_percentile_P2_SpectrumSharing_BS128_UE1} that most UEs suffer because of spectrum sharing. Full bandwidth sharing leads to the worst performance for the 5th percentile UEs and for the 50th percentile UEs, which is due to high inter-operator interference.
In Fig.~\ref{subfig: Rate_percentile_P2_SpectrumSharing_BS1024_UE128}, we repeat the previous comparison under the assumption of a sufficiently large number of antennas both at the UEs and at the BSs. In this case, both partial and full sharing enhance the 5th, 50th, and 95th percentiles compared to the baseline (i.e., exclusive). This figure suggests that spectrum sharing may be harmful for signals that are transmitted/received with large beamwidths. Examples of these signals include broadcast control messages. Since these signals usually have low SINR due to the lack of antenna gains, they should be protected from the adverse effect of spectrum sharing, e.g., by exclusive resource allocation for those critical signals.

\begin{figure}[t]
  \centering
  \includegraphics[width=\columnwidth]{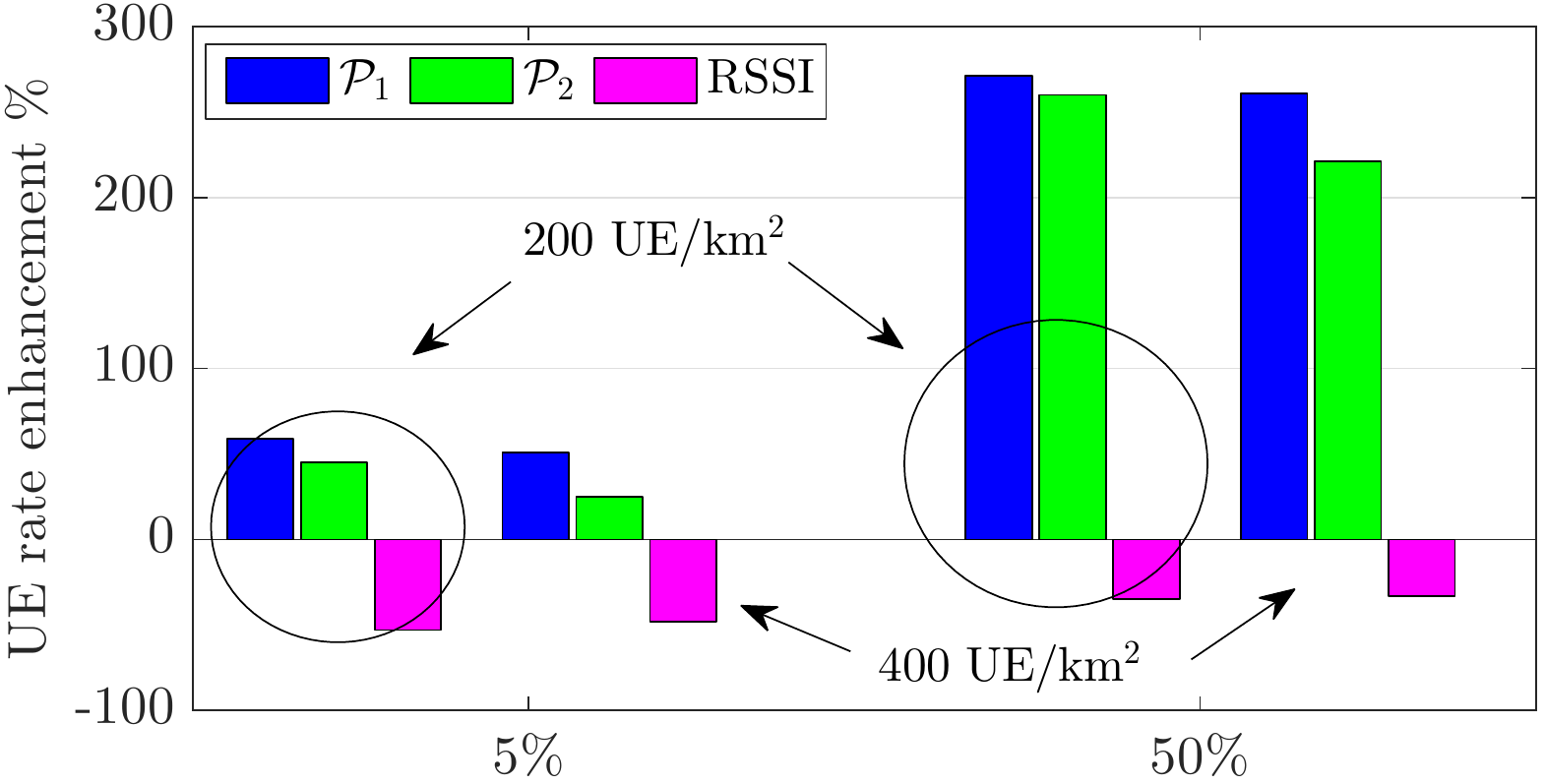}

  \caption{Full bandwidth sharing performance, with/without inter-operator coordination, assuming analog precoding with $N_{\mathrm{BS}} = 256$ and $N_{\mathrm{UE}} = 16$ and a carrier frequency of 32 GHz.}
  \label{fig: Rate_percentie_UEdensity}
\end{figure}
Fig.~\ref{fig: Rate_percentie_UEdensity} shows the impact of the user density on the spectrum sharing performance, under different information sharing scenarios. The baseline is an exclusive spectrum allocation with optimal association ($\mathcal{P}_3$) for a mmWave network with 100 BSs/km$^2$. From the perspective of a serving BS, increasing the UE density reduces the angular separation among its UEs. Therefore, serving one of them with an analog beam will cause higher interference to its other UEs in non-asymptotic regimes,\footnote{Note that the interference terms are zero almost surely in the asymptotic regime; see Remark~\ref{remark: asymptotic-no-int}.} compared to the interference level in a sparser network. Such higher multiuser interference reduces the benefits of spectrum sharing, which is more prominent in the scenarios with less coordination ($\mathcal{P}_2$ and RSSI) due to the major role of coordination in interference reduction and load balancing; see Fig.~\ref{subfig: Interference_average}.

\begin{figure}[!t]
  \centering
  \begin{subfigure}[t]{\columnwidth}
    \centering
    \includegraphics[width=\columnwidth]{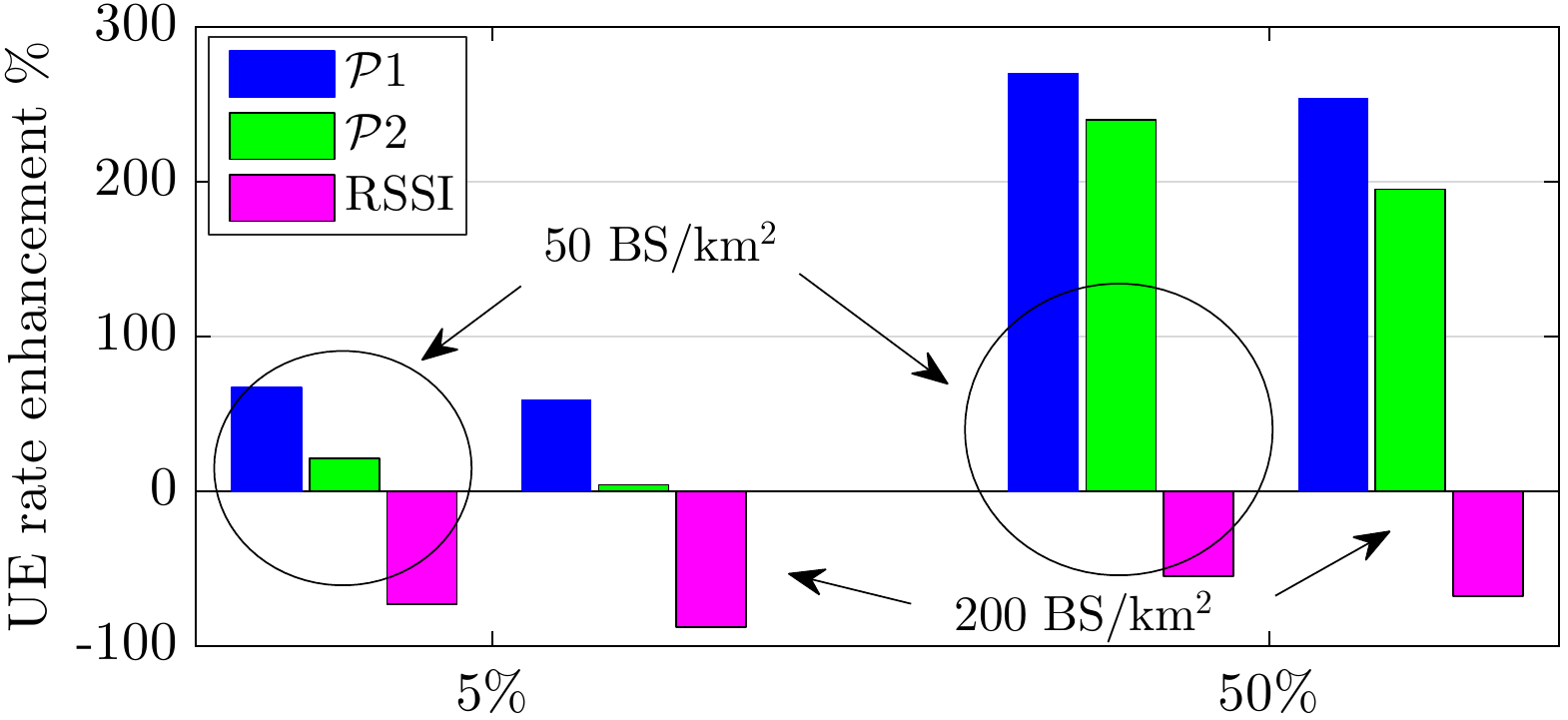}
    \caption{32 GHz}\label{subfig: Rate_percentile_P1_P2_P3_BS128_UE8}
  \end{subfigure}
  \begin{subfigure}[t]{\columnwidth}
       \centering
	\includegraphics[width=\columnwidth]{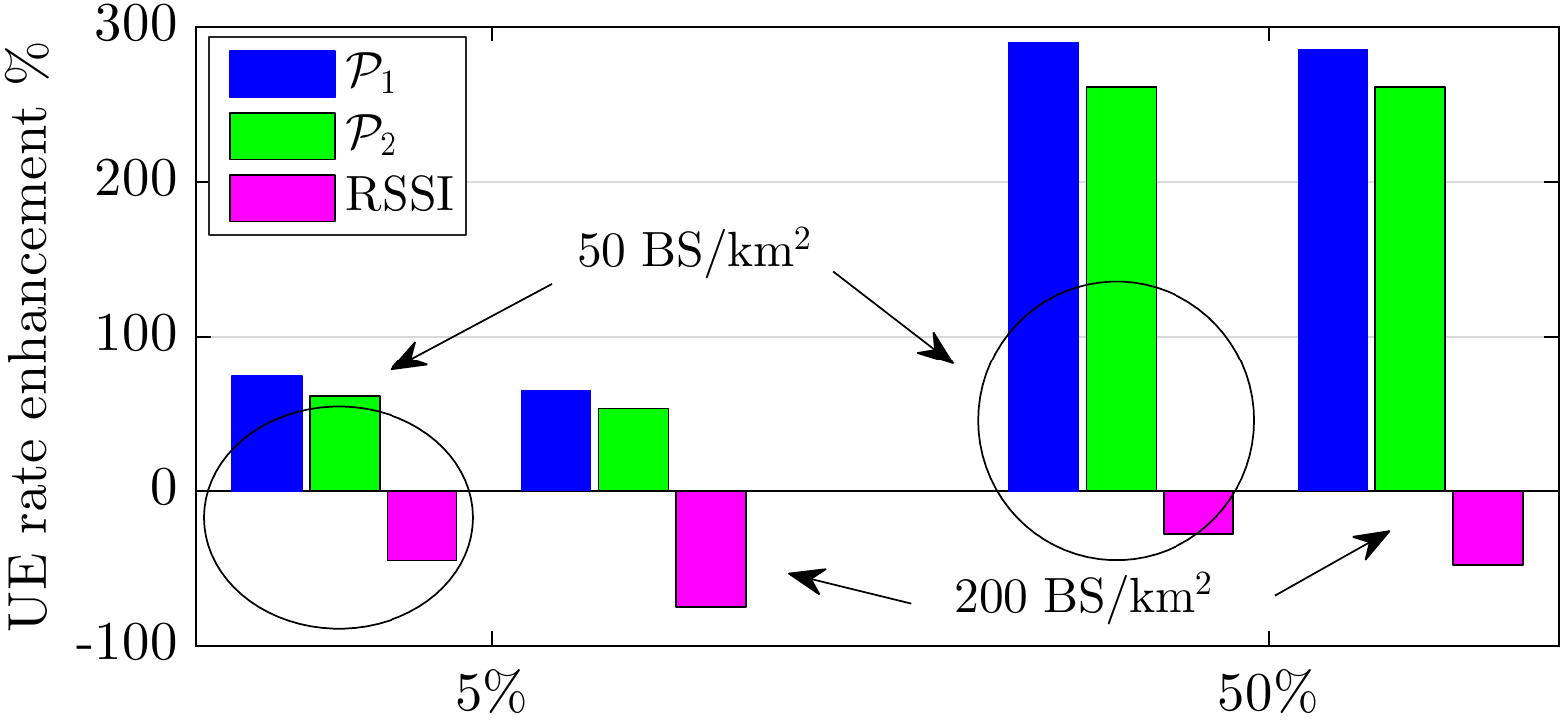}
    \caption{73 GHz}\label{subfig: Rate_percentile_P1_P2_P3_BS1024_UE128}
    \end{subfigure}

  \caption{Full bandwidth sharing performance, with/without inter-operator coordination, assuming analog precoding with $N_{\mathrm{BS}} = 256$ and $N_{\mathrm{UE}} = 16$.}
  \label{fig: Rate_percentile_P1_P2_P3}
\vspace{-2mm}
\end{figure}
Fig.~\ref{fig: Rate_percentile_P1_P2_P3} shows the impact of the operating frequency and the BS density on the spectrum sharing performance, under different information sharing scenarios. We consider transmissions at 32~GHz and 73~GHz. We keep the dimension of the antenna array constant as a function of the frequency, i.e., at 73 GHz we consider twice as many antenna elements as at 32 GHz, at both BSs and UEs. The baseline is an exclusive spectrum allocation with optimal association ($\mathcal{P}_3$), for different values of the BS density: 50 and 200 BSs/km$^2$ (corresponding to a cell radius of 80~m and 39~m, respectively).
We note that, similar to increasing the user density, increasing the BS density of individual operators exacerbates the multiuser interference. Again, lack of coordination may substantially degrade the performance of spectrum sharing. At 73~GHz, larger antenna arrays help form narrower beams, which in turn leads to less multiuser interference. Therefore, a mmWave communication at 73~GHz is less sensitive to inter-operator coordination, which is only for the sake of interference rejection.
However, intra-operator coordination is still beneficial, mostly because it enables load balancing within an operator. The UE rate performance of a network with RSSI-based association suffers from some highly loaded BSs, in addition to intra- and inter-cell interference.

\subsection{Digital Precoding}
\begin{figure}[!t]
  \centering
    \includegraphics[width=\columnwidth]{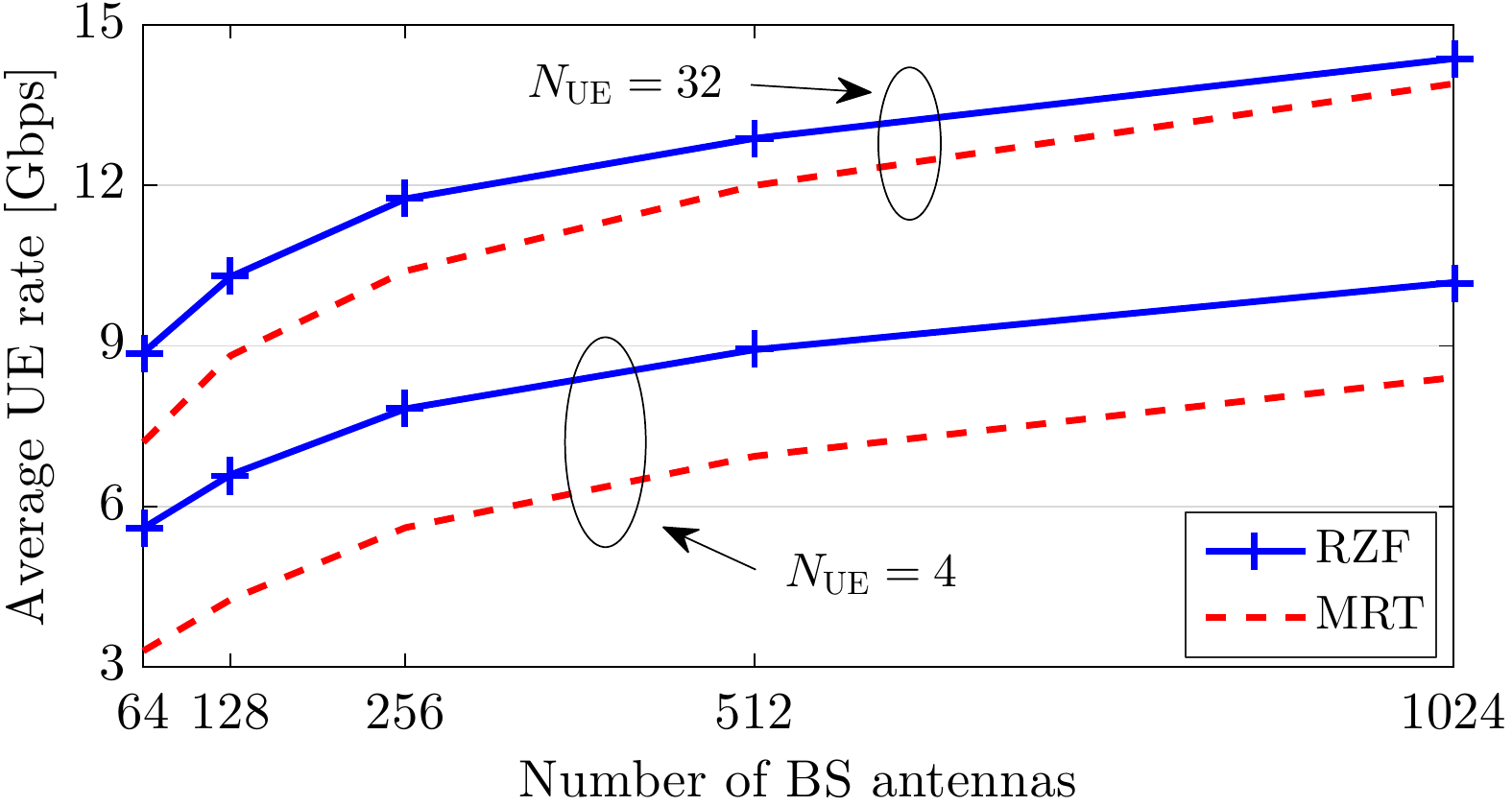}
    \caption{Performance of the full bandwidth and information sharing scenario with digital precoders ($\mathcal{P}_4$).}
    \label{fig: Rate_average_P1_Digital}
\end{figure}
Assuming full bandwidth and information sharing ($\mathcal{P}_4$), Fig.~\ref{fig: Rate_average_P1_Digital} compares the performance of spectrum sharing when using different precoding schemes. Generally, the average rate of the UEs increases with the number of antenna elements in both precoding strategies. Particularly, in the RZF precoder, this rate enhancement is mainly due to higher antenna gains. In the MRT precoder, it is primarily due to both higher antenna gains and lower multiuser interference. Again, UE antennas have a complementary effect to the number of BS antennas, and a few more antennas at the UEs can maintain the same performance as adding many antennas at the BSs. Moreover, the RZF precoder outperforms the MRT precoder, as it completely cancels all the interference components. The performance gap, however, vanishes when either $N_{\mathrm{BS}}$ or $N_{\mathrm{UE}}$ grows large; see also Proposition~\ref{prop: Asymptotic-digital2}.

\begin{figure}[t]
    \centering
	\includegraphics[width=\columnwidth]{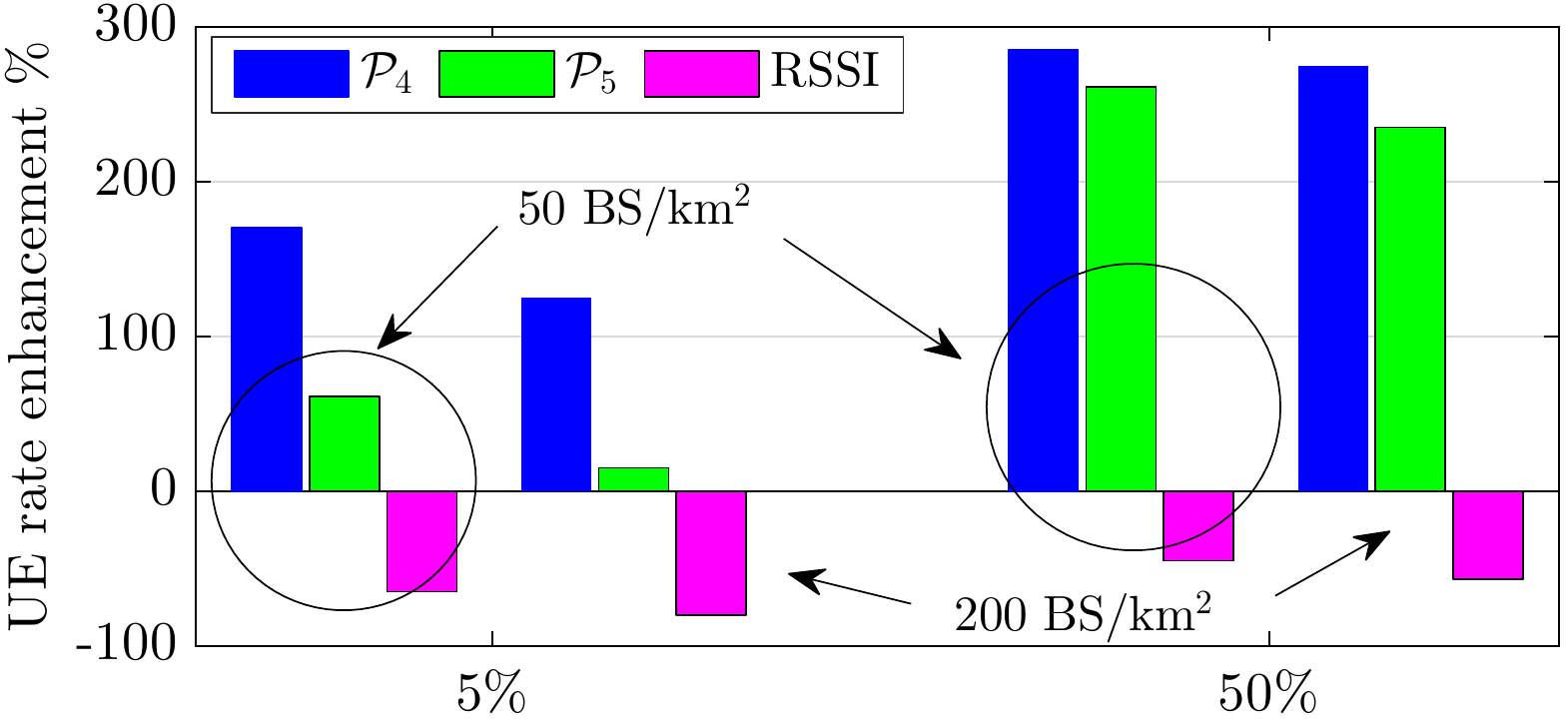}
      \caption{Full bandwidth sharing performance, with/without inter-operator coordination, assuming RZF precoding with $N_{\mathrm{BS}} = 256$ and $N_{\mathrm{UE}} = 16$.}
\label{fig: Rate_percentile_P1_P2_P3_BS128_UE8_Digital}
\end{figure}
Fig.~\ref{fig: Rate_percentile_P1_P2_P3_BS128_UE8_Digital} illustrates the effect of inter-operator coordination and BS density on the 5th and the 50th UE rates percentiles, assuming RZF precoding. The baseline is the UE rate achieved by the optimal association with exclusive bandwidth allocation $\mathcal{P}_5$. The benefits of using digital precoding with full information sharing ($\mathcal{P}_4$) is evident in the 5th percentile of the UE rates when compared to that in Fig.~\ref{subfig: Rate_percentile_P1_P2_P3_BS128_UE8}. The main reason is that the RZF precoder can completely cancel all the interference terms, which protects the weakest UEs (5th percentile) from inter-operator interference after spectrum sharing. This protection becomes weaker when using analog precoding (in the non-asymptotic regime). Nonetheless, we note that the design of the optimal analog precoder may need only second-order statistics of the channel (Remark~\ref{remark: Single-path-ABF}), which imposes substantially less overhead on the channel estimation procedure. Higher BS density corresponds to stronger multiuser interference components, which can be entirely canceled in $\mathcal{P}_4$. In $\mathcal{P}_4$, we cannot cancel inter-operator interference, leading to a significant performance drop in the 5th percentile compared to $\mathcal{P}_5$. This performance drop between $\mathcal{P}_4$ and $\mathcal{P}_5$ reduces as the SINR of the UEs increases (50th percentile). Again, the RSSI-based association approach has the poorest performance and can lead to some overloaded BSs, and therefore very low UE rates.


\section{Design Insights}\label{sec: design-insights}
In previous sections, we have shown that under the assumption of ideal channel estimation, both analog and digital precoding schemes can realize spectrum sharing, even though digital precoder has superior performance. Larger antenna arrays either at the BSs or at the UEs improve the performance of spectrum sharing  by reducing the interference components. Coordination has two effects: 1) balancing the network load within an operator, and 2) reducing the intra- and inter-operator interference. As the number of antennas grows large, the importance of inter-operator coordination vanishes, as large antenna arrays can cancel interference. Still, intra-operator coordination is beneficial due to the load balancing gain. In the non-asymptotic regime, the existence of some sporadic interference necessitates on-demand coordination, which imposes much less overhead on the core network compared to the coordination required in sub-6~GHz networks to achieve similar network performance gains. Moreover, coordination is more critical at 32~GHz than at 73~GHz, due to the fact that at lower frequencies beamforming by itself is not sufficient to protect the weakest users from inter-operator interference. Last but not least, we highlight that critical control signals should be protected from the adverse effects of spectrum sharing, e.g., by exclusive resource allocation. Table~\ref{table: summary} summarizes the main conclusions reached in the paper.
\begin{table}[t]
  \centering
  \caption{Summary of the main conclusions reached in the paper.}\label{table: summary}
{
\renewcommand{\arraystretch}{1.5}
  {
\begin{tabular}{|m{0.6\columnwidth}|m{0.3\columnwidth}|}
\hline
   \textbf{Conclusions} & \textbf{Supported by} \\ \hline
   Directional communications at the UEs substantially alleviate the disadvantages of spectrum sharing (e.g., higher multiuser interference). & Remarks~\ref{remark: asymptotic-no-int} and~\ref{remark: SameSolution}, Proposition~\ref{prop: Asymptotic-digital2}, and Figs.~\ref{fig: AveragePerformanceP1}--\ref{fig: Rate_percentile_P2} \\ \hline
   Larger antenna sizes can reduce the need for coordination and simplify the realization of spectrum sharing. & Remarks~\ref{remark: asymptotic-no-int} and~\ref{remark: SameSolution}, Proposition~\ref{prop: Asymptotic-digital2}, and Figs.~\ref{fig: AveragePerformanceP1}--\ref{fig: Rate_percentile_P2} \\ \hline
   Inter-operator coordination can be neglected in the large antenna regime; however, intra-operator coordination can still bring gains by balancing the network load. & Figs.~\ref{fig: Rate_percentile_P1_P2_P3} and~\ref{fig: Rate_percentile_P1_P2_P3_BS128_UE8_Digital} \\ \hline
   Spectrum sharing with light on-demand intra- and inter-operator coordination is promising in mmWave networks, especially at higher mmWave frequencies (e.g., 73 GHz). & Fig.~\ref{fig: Rate_percentile_P1_P2_P3} \\ \hline
   Critical control signals should be protected from the adverse effects of spectrum sharing, e.g., by exclusive resource allocation. & Fig.~\ref{fig: Rate_percentile_P2} \\ \hline
\end{tabular}
}}
\end{table}

Although the idea of spectrum sharing in mmWave networks is similar to that in traditional wireless networks, the way we should realize it and the achievable performance at mmWave are significantly different than in those networks. The main reasons are \emph{i}) sparsity of the mmWave channel in the angular domain (less possibility for unintended transmitters to create interference) as opposed to rich scattering characteristics of lower frequencies, \emph{ii}) blockage, and \emph{iii}) directional transmissions and receptions. Another fundamental difference is that mmWave communications will take advantage of its bandwidth being orders of magnitude wider than in the sub-6~GHz bands. Due to the unique features of mmWave networks, the role of coordination is substantially more prominent in sub-6~GHz networks compared to mmWave networks. This indeed means that we need substantially different protocol and architectural enablers (with lighter overhead) for spectrum sharing at mmWave compared to that of traditional networks.

From Table~\ref{table: summary}, a reliable control channel for exchanging coordination information is a key enabler of spectrum sharing in mmWave networks. Indeed, there is a tradeoff between performance enhancement by coordination (e.g., interference reduction) and the corresponding complexity (to guarantee delay and reliability requirements) of the control plane. For example, in backhaul networks, the multiuser interference level is generally so low (due to small user density and high number of antennas) that neglecting inter-operator coordination leads to almost no performance loss.  For an ultra-dense mmWave network, however, coordination may bring significant gains, especially for the weakest (5th percentile) UEs, as shown in Fig.~\ref{fig: Rate_percentile_P1_P2_P3}. These gains become available only by a reliable control channel. In~\cite{shokri2015mmWavecellular}, several mechanisms to realize a reliable control channel in mmWave cellular networks are presented.

This paper was ultimately motivated by the need to understand the performance gains of spectrum sharing in mmWave communications and the benefits of beamforming and coordination.
Our results suggest that, unlike in current mobile networks, spectrum sharing between mmWave mobile networks using beamforming has a potential to work well and improve spectral efficiency. We believe that these results, as well as further research that this work will stimulate in this area, could provide some new and relevant insights to standardization and spectrum policy bodies addressing the future regulation of mmWave bands.

\section{Conclusions}\label{sec: concluding-remarks}
Sharing the spectrum among multiple operators can provide larger bandwidth to individual UEs at the expense of increased inter-operator interference. The typical characteristics of mmWave systems, such as high penetration loss and directional communications both at the transmitter and at the receiver, may overcome such problem and potentially enable large benefits. In this paper, we have investigated the extent of these benefits by proposing an optimization framework based on a joint beamforming design and BS association, with the objective of maximizing the long-term throughput of the UEs. Specifically, we have developed an optimization framework that allows quantifying the gains of (1) beamforming, (2) coordination, (3) bandwidth sharing, and (4) sharing architecture. We have analyzed the gains of these parameters in the asymptotic and non-asymptotic regimes, and showed that  (1) directional communications at the UEs substantially alleviate the potential disadvantages of spectrum sharing (such as higher multiuser interference); (2) larger antenna sizes can reduce the need for coordination and simplify the realization of spectrum sharing; (3) intra-operator coordination can always bring gains by balancing the network load, whereas the gain of inter-operator coordination vanishes in the large-antenna regime; (4) spectrum sharing with light on-demand intra- and inter-operator coordination is promising in mmWave networks, especially at higher mmWave frequencies (such as 73 GHz); and finally (5) critical control signals should be protected from the adverse effects of spectrum sharing, for example by means of exclusive resource allocation.

Although the focus of this paper was about stand-alone mmWave wireless networks, the proposed framework can be naturally used to model various kinds of spectrum sharing scenarios such as coexistence of sub-6~GHz and mmWave networks, of backhaul and access networks, and of cellular and device-to-device networks.
Future work will be devoted to investigating non-ideal assumptions on spectrum sharing, such as pilot contamination, imperfect channel estimation, UE mobility, and distributed beamforming and association methods. Optimizing the power allocation per individual RF chains in the analog precoding and per individual UEs in digital precoding is another future direction.
Moreover, although spectrum sharing seems feasible (from a technical perspective) in most mmWave network settings, its practicality may depend on non-technical issues such as the desirability of promoting competition, encouraging investments and innovation, and achieving a widespread availability of services across rural and urban areas. Further work is needed to further study these and other non-technical factors as well.



\bibliographystyle{IEEEtran}
\bibliography{References}

\end{document}